\documentclass[11pt, letterpaper, onecolumn]{IEEEtran}

\usepackage[mathscr]{eucal}
\usepackage[cmex10]{amsmath}
\usepackage{epsfig,epsf,psfrag}
\usepackage{amssymb,amsmath,amsthm,amsfonts,latexsym}
\usepackage{amsmath,graphicx,bm,xcolor,url}
\usepackage[caption=false]{subfig} 
\usepackage{fixltx2e}
\usepackage{array}
\usepackage{verbatim}
\usepackage{bm}
\usepackage{algorithmic, cite}
\usepackage{algorithm}
\usepackage{verbatim}
\usepackage{textcomp}
\usepackage{mathrsfs}
\usepackage{epstopdf}

\catcode`~=11 \def\UrlSpecials{\do\~{\kern -.15em\lower .7ex\hbox{~}\kern .04em}} \catcode`~=13 

\allowdisplaybreaks[4]
 
\newcommand{\nn}{\nonumber}

\newcommand{\calA}{\mathcal{A}}
\newcommand{\calB}{\mathcal{B}}

\newcommand{\calD}{\mathcal{D}}
\newcommand{\calE}{\mathcal{E}}
\newcommand{\calF}{\mathcal{F}}

\newcommand{\calM}{\mathcal{M}}

\newcommand{\calS}{\mathcal{S}}
\newcommand{\calT}{\mathcal{T}}
\newcommand{\calU}{\mathcal{U}}

\newcommand{\calX}{\mathcal{X}}
\newcommand{\calY}{\mathcal{Y}}
\newcommand{\calZ}{\mathcal{Z}}

\newcommand{\boldm}{\mathbf{m}}

\newcommand{\rmC}{\mathrm{C}}

\newcommand{\rmD}{\mathrm{D}}
\newcommand{\rme}{\mathrm{e}}

\newcommand{\rmH}{\mathrm{H}}

\newcommand{\rmP}{\mathrm{P}}

\newcommand{\rmr}{\mathrm{r}}
\newcommand{\rmR}{\mathrm{R}}

\newcommand{\bbE}{\mathbb{E}}

\newcommand{\bbN}{\mathbb{N}}

\newcommand{\bbP}{\mathbb{P}}

\newcommand{\bbR}{\mathbb{R}}

\newcommand{\scG}{\mathscr{G}}

\newcommand{\scK}{\mathscr{K}}

\newcommand{\scP}{\mathscr{P}}

\newcommand{\scR}{\mathscr{R}}

\newcommand{\scV}{\mathscr{V}}

\DeclareMathAlphabet{\mathbsf}{OT1}{cmss}{bx}{n}
\DeclareMathAlphabet{\mathssf}{OT1}{cmss}{m}{sl}

\newcommand{\rvC}{\mathsf{C}}

\DeclareSymbolFont{bsfletters}{OT1}{cmss}{bx}{n}  
\DeclareSymbolFont{ssfletters}{OT1}{cmss}{m}{n}
\DeclareMathSymbol{\bsfGamma}{0}{bsfletters}{'000}
\DeclareMathSymbol{\ssfGamma}{0}{ssfletters}{'000}
\DeclareMathSymbol{\bsfDelta}{0}{bsfletters}{'001}
\DeclareMathSymbol{\ssfDelta}{0}{ssfletters}{'001}
\DeclareMathSymbol{\bsfTheta}{0}{bsfletters}{'002}
\DeclareMathSymbol{\ssfTheta}{0}{ssfletters}{'002}
\DeclareMathSymbol{\bsfLambda}{0}{bsfletters}{'003}
\DeclareMathSymbol{\ssfLambda}{0}{ssfletters}{'003}
\DeclareMathSymbol{\bsfXi}{0}{bsfletters}{'004}
\DeclareMathSymbol{\ssfXi}{0}{ssfletters}{'004}
\DeclareMathSymbol{\bsfPi}{0}{bsfletters}{'005}
\DeclareMathSymbol{\ssfPi}{0}{ssfletters}{'005}
\DeclareMathSymbol{\bsfSigma}{0}{bsfletters}{'006}
\DeclareMathSymbol{\ssfSigma}{0}{ssfletters}{'006}
\DeclareMathSymbol{\bsfUpsilon}{0}{bsfletters}{'007}
\DeclareMathSymbol{\ssfUpsilon}{0}{ssfletters}{'007}
\DeclareMathSymbol{\bsfPhi}{0}{bsfletters}{'010}
\DeclareMathSymbol{\ssfPhi}{0}{ssfletters}{'010}
\DeclareMathSymbol{\bsfPsi}{0}{bsfletters}{'011}
\DeclareMathSymbol{\ssfPsi}{0}{ssfletters}{'011}
\DeclareMathSymbol{\bsfOmega}{0}{bsfletters}{'012}
\DeclareMathSymbol{\ssfOmega}{0}{ssfletters}{'012}

\newcommand{\tilE}{\tilde{E}}

\newcommand{\tilG}{\tilde{G}}

\newcommand{\hatI}{\hat{I}}

\newcommand{\hatk}{\hat{k}}

\newcommand{\tilk}{\tilde{k}}

\newcommand{\hatl}{\hat{l}}
\newcommand{\hatL}{\hat{L}}
\newcommand{\till}{\tilde{l}}

\newcommand{\hatm}{\hat{m}}
\newcommand{\hatM}{\hat{M}}
\newcommand{\tilm}{\tilde{m}}

\newcommand{\hatQ}{\hat{Q}}

\newcommand{\tilQ}{\tilde{Q}}

\newcommand{\tilR}{\tilde{R}}

\newcommand{\tilU}{\tilde{U}}

\newcommand{\hatV}{\hat{V}}

\newcommand{\tilV}{\tilde{V}}

\newcommand{\tilX}{\tilde{X}}

\newcommand{\haty}{\hat{y}}
\newcommand{\hatY}{\hat{Y}}

\newcommand{\tilY}{\tilde{Y}}

\newcommand{\barx}{\bar{x}}
\newcommand{\bary}{\bar{y}}

\newcommand{\barW}{\bar{W}}
\newcommand{\barX}{\bar{X}}

\newcommand{\btheta}{\bm{\theta}}

\newcommand{\veps}{\varepsilon}

\newcommand{\bTheta}{\bm{\Theta}}

\def\fndot{\, \cdot \,}

\newcommand{\ceil}[1]{\lceil{#1}\rceil}

\newcommand{\dotleq}{\stackrel{.}{\leq}}

\DeclareMathOperator*{\argmax}{arg\,max}
\DeclareMathOperator*{\argmin}{arg\,min}

\newcommand{\Unif}{\mathrm{Unif}}

\newcommand{\bone}{\mathbf{1}}

\newtheorem{theorem}{Theorem} 
\newtheorem{lemma}{Lemma}

\newtheorem{definition}{Definition}

\newcommand{\qednew}{\nobreak \ifvmode \relax \else
      \ifdim\lastskip<1.5em \hskip-\lastskip
      \hskip1.5em plus0em minus0.5em \fi \nobreak
      \vrule height0.75em width0.5em depth0.25em\fi}

\newcommand{\Reff}{R_{\mathrm{eff}}}
\newcommand{\hcalY}{\hat{\calY}}

\newcommand{\cp}{\!\times \!}

\usepackage{cite}
\allowdisplaybreaks[1]

\title{On the Reliability Function of the \\ Discrete Memoryless Relay Channel}
\author{Vincent Y.~F.\ Tan, {\em Member, IEEE} \thanks{The author is with the Department of Electrical and Computer Engineering (ECE) and the Department of Mathematics at the  National University of Singapore (Email: vtan@nus.edu.sg).   }\thanks{This paper was presented in part at the 2013 International Symposium on Information Theory in Istanbul, Turkey. }\thanks{The work of the author is supported in part by NUS startup grant R-263-000-A98-750/133 and in part by A*STAR, Singapore. } } 

\begin{document}
\flushbottom
\maketitle

\begin{abstract} 
Bounds on the reliability function for the discrete memoryless relay channel are derived using the method of types. Two achievable  error exponents are derived based on  partial decode-forward and compress-forward which are well-known superposition block-Markov coding schemes.  The  derivations   require  combinations of the techniques involved in the proofs of   Csisz\'{a}r-K\"{o}rner-Marton's packing lemma for  the error exponent of channel coding  and Marton's  type covering lemma for   the error exponent of   source coding with a fidelity criterion. The decode-forward error exponent is evaluated on Sato's relay channel. From this example, it is noted that to obtain the fastest possible decay in the error probability for a fixed effective coding rate, one ought to optimize the number of blocks in the block-Markov coding scheme assuming the blocklength within each block is   large.  An upper bound on the reliability function is also derived using ideas from Haroutunian's lower bound on the error probability for point-to-point channel coding with   feedback. 
\end{abstract}

\section{Introduction}\label{sec:intro}
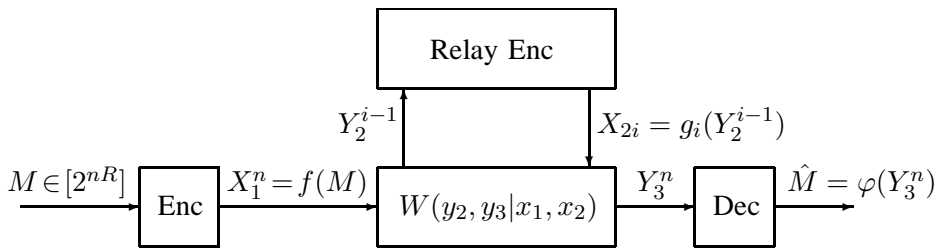
\begin{figure}
\centering
\begin{picture}(310,90)
\linethickness{.25mm}
\put(30,0){\line(1,0){30}}
\put(30,0){\line(0,1){30}}
\put(60,0){\line(0,1){30}}
\put(30,30){\line(1,0){30}}


\put(120,0){\line(1,0){90}}
\put(120,0){\line(0,1){30}}
\put(210,0){\line(0,1){30}}
\put(120,30){\line(1,0){90}}

\put(240,0){\line(1,0){30}}
\put(240,0){\line(0,1){30}}
\put(270,0){\line(0,1){30}}
\put(240,30){\line(1,0){30}}


\put(-15,15){\vector(1,0){45}}
\put(60,15){\vector(1,0){60}}
\put(210,15){\vector(1,0){30}}
\put(270,15){\vector(1,0){30}}

\put(120,60){\line(1,0){90}}
\put(120,60){\line(0,1){30}}
\put(210,60){\line(0,1){30}}
\put(120,90){\line(1,0){90}}

\put(130,30){\vector(0,1){30}}
\put(200,60){\vector(0,-1){30}}

\put(140,71){\mbox{Relay Enc}}


\put(128,12){\mbox{$W(y_2, y_3|x_1 , x_2)$}}
\put(247,12){\mbox{Dec}}
\put(37,12){\mbox{Enc}}

\put(-20,21){\mbox{$M\! \in \! [2^{nR}]$}}
\put(275,21){\mbox{$\hat{M} = \varphi(Y_3^n)$}}
\put(218,21){\mbox{$Y_3^n$}}
\put(63,21){\mbox{$X_1^n \!=\! f(M)$}}


\put(105,43){\mbox{$Y_2^{i-1}$}}
\put(203,43){\mbox{$X_{2i}=g_i(Y_2^{i-1})$}}

\end{picture}
\caption{The relay channel with the notations we use in this paper}
\label{fig:relay}
\end{figure}

We derive bounds on the reliability function for the discrete memoryless relay channel. This channel, introduced  by van der Meulen in~\cite{vdM}, is a point-to-point communication system that consists of  a sender $X_1$, a receiver $Y_3$  and a relay  with input $Y_2$ and output $X_2$. See Fig.~\ref{fig:relay}. The capacity is not known in general but there exist  several coding schemes that are optimal for certain classes of relay channels, e.g., physically degraded. These coding schemes, introduced in the seminal work by Cover and El Gamal~\cite{CEG} include decode-forward (DF), partial decode-forward (PDF) and compress-forward (CF). Using PDF, the capacity of the relay channel  $C$ is lower bounded as 
\begin{align}
C \ge  \max  \min\{ I(U X_2; Y_3),    I(U ;Y_2|X_2)   +   I(X_1;Y_3|X_2  U)\} \label{eqn:cap_pdf}
\end{align}
where the maximization is over all   $P_{U X_1  X_2}$. The auxiliary random variable $U$ with cardinality  $|\calU|\le |\calX_1| |\calX_2|$ represents a part of the message that the relay decodes; the rest of the messsage is decoded by the receiver. DF is a special case of PDF in which $U=X_1$ and instead of decoding part of the message as in PDF, the relay  decodes the entire message.  In CF, a more complicated coding scheme, the relay sends a description of $Y_2$ to the receiver. This description is denoted as $\hatY_2$.  The receiver then uses $Y_3$ as side information \`a la Wyner-Ziv~\cite[Ch.\ 11]{elgamal} \cite{wynerziv} to reduce the rate of the description. One form of the CF lower bound is given as~\cite{EMZ06} 
\begin{align}
C\ge \max \min\{ I(X_1;\hatY_2  Y_3|X_2),I(X_1  X_2;Y_3)  -I(Y_2;\hatY_2|X_1  X_2 Y_3)\}\label{eqn:cf}
\end{align}
where the maximization is over     $P_{X_1}, P_{X_2}$ and $P_{\hatY_2|X_2  Y_2}$ and $|\hcalY_2|\le |\calX_2| |\calY_2| + 1$.  Both PDF and CF involve {\em  block-Markov coding}~\cite{CEG} in which the channel is used $N=nb$ times over $b$   blocks, each involving an independent message to be sent and the relay codeword in block $j$  depends statistically on the message from block $j-1$.  The best known upper bound on the capacity is the so-called cutset bound~\cite{CEG} 
\begin{align}
C\le\max \min\{ I(X_1 X_2; Y_3), I(X_1; Y_2 Y_3|X_2)\} \label{eqn:cutset_intro}
\end{align}
where the maximization is over all $P_{X_1 X_2}$. 

In addition to capacities, in information theory, error exponents are also of tremendous interest. They quantify the exponential rate of decay of the error probability when the  rate of the code is below capacity or the set of rates is strictly within the capacity region.  Such results allow us to provide approximate bounds on the blocklength needed to achieve a certain rate (or set of rates) and so provide a means to understand the tradeoff between rate(s) and error probability. In this paper, we derive achievable  error exponents based on two superposition block-Markov coding schemes for the   discrete memoryless relay channel. These bounds are positive for all rates below \eqref{eqn:cap_pdf} and \eqref{eqn:cf}. We also derive an upper bound on the reliability function that is positive for all rates below the cutset bound in \eqref{eqn:cutset_intro}. We evaluate the   exponent based on PDF    for the Sato relay channel~\cite{sato76}.

\subsection{ Main Contributions} \label{sec:main_contr}
We  now elaborate on our three main contributions in this paper--all concerning bounds on the reliability function.

For  PDF, of which DF is a special case,  by using maximum mutual information (MMI) decoding~\cite{Goppa,Csi97}, we show that the analogue of the random coding error exponent (i.e., an error exponent that is similar in style to the one presented in~\cite[Thm.~10.2]{Csi97}) is {\em universally attainable}. That is, the decoder does not need to know the channel statistics.  This is in contrast to the recent work by Bradford-Laneman~\cite{Bra12} in which the authors employed maximum-likelihood  (ML) decoding with the sliding window decoding technique introduced by Carleial~\cite{Carleial} and also used by Kramer-Gastpar-Gupta~\cite{KGG05} for relay networks. In~\cite{Bra12}, the channel needs to be known at the decoder but  one advantage the sliding window has over backward decoding~\cite{Zeng89, Willems85}   (which we use) is that it ameliorates the problem of excessive delay. To prove this result, we generalize the techniques used to prove the packing lemmas in~\cite{Csi97, Csis00, Kor80b, CKM, Har08} so that they are  applicable to the relay channel. 

For CF, we draw inspiration from~\cite{Kel12}  in which the authors derived achievable error exponents for Wyner-Ahlswede-K\"orner coding (lossless source coding with coded side information)~\cite[Ch.~10]{elgamal}  and Wyner-Ziv (lossy source coding with decoder side information)~\cite[Ch.~11]{elgamal}  coding. We  handle the combination of covering and packing  in a similar way as~\cite{Kel12} in order to derive an achievable error exponent for CF. In addition, a key technical contribution is the taking into account of the conditional correlation between $\hatY_2$ and $X_1$ (given $X_2$) using a bounding technique introduced by Scarlett-Martinez-Guill\'en i F\`abregas~\cite{Sca12, Sca13Nov} called, in our words, the {\em one-at-a-time union bound}. This bound is reviewed in Section~\ref{sec:oneat}. 
We also leverage   Csisz\'ar's $\alpha$-decoder~\cite{CK81}  which specializes, in the point-to-point case, to ML decoding~\cite[Ch.~5]{gallagerIT} and MMI decoding~\cite{Goppa}.

For the upper bound on the reliability function, we draw on ideas from Haroutunian's lower bound on the error probability for channel coding with  feedback~\cite{Har77}.  Our proof leverages on work by Palaiyanur~\cite{Palai11}.  We show that the upper bound can be expressed similarly to Haroutunian's bound with the constraint that the minimization is over all transition matrices for which the cutset bound is no larger than the rate of transmission. This is the first time an upper bound  on the reliability function for relay channels has been derived. At a very high level, we cast the relay channel as a  point-to-point channel from $(X_1, X_2)$ to $(Y_2, Y_3)$ with feedback and make use of techniques developed by Haroutunian~\cite{Haroutunian68, Har77} and Palaiyanur~\cite{Palai11}. 
\subsection{Related Work} \label{sec:related}
The work that is most closely related to the current one are the papers by Bradford-Laneman~\cite{Bra12}  and Nguyen-Rasmussen~\cite{nguyen13} who derived  random coding error exponents for DF based on Gallager's Chernoff-bounding techniques~\cite{gallagerIT}. The latter paper considers a streaming setting as well. We generalize their  results to PDF and  we use MMI decoding which has the advantage of being universal. Our techniques, in contrast to all previous works on error exponents for relaying, hinge on the method of types which we find convenient in the  discrete memoryless setting. The results are also intuitive and can be interpreted easily.  For PDF, our work leverages on techniques used to prove various forms of the packing lemmas for multiuser channels in, for example, in the monograph by Haroutunian-Haroutunian-Harutyunyan~\cite{Har08}. It also uses a change-of-measure technique  introduced by Hayashi~\cite{Hayashi09} for proving second-order coding rates in channel coding. This change-of-measure technique allows us to use a random  constant composition code ensemble  and subsequently analyze this ensemble as if it were an i.i.d.\ ensemble without any loss in the error exponents sense. 

For CF, since it is closely related to Wyner-Ziv coding~\cite{wynerziv}, we leverage on the work of Kelly-Wagner~\cite{Kel12} who derived an achievable exponents for  Wyner-Ahlswede-K\"orner coding~\cite[Ch.~10]{elgamal}  and Wyner-Ziv~\cite[Ch.~11]{elgamal}  coding. In a similar vein,  Moulin-Wang~\cite{Mou07} and Dasarathy-Draper~\cite{Dar11}  derived lower bounds for the error exponents of Gel'fand-Pinsker coding~\cite[Ch.~7]{elgamal}  and content identification respectively. These works involve analyzing the combination of both packing and covering error events. 

We also note that the authors in~\cite{Ngo10}  and \cite{Yilmaz} presented achievable error exponents for various schemes  for the additive white Gaussian noise (AWGN) relay channel and backhaul-constrained parallel relay networks respectively but these works do not take into account block-Markov coding~\cite{CEG}. Similarly, \cite{LiGeor} analyzes the error exponents for fading Gaussian relay channels but  does not take into account block-Markov coding.  There is also a collection of closely-related works addressing error exponents of multihop networks~\cite{zhangmitra, WenBerry} as the number of blocks used affects both reliability and rate of a given block-Markov scheme. We study this effect in detail for the decode-forward scheme applied to the  Sato relay channel.


\subsection{Structure of Paper}
This paper is structured as follows. In Section~\ref{sec:prelim}, we state our notation, some standard results from the method of types~\cite{Csi97, Csis00} and the definitions of the discrete memoryless relay channel, and the reliability function. To make this paper as self-contained as possible, prior reliability function results for the point-to-point channel and for lossy source coding are also reviewed   in this section. In Sections~\ref{sec:dec} and \ref{sec:cf}, we state and prove error exponent theorems for PDF and CF respectively.  In Section~\ref{sec:sp} we state an upper bound on the reliability function. In  these three technical sections, the proofs of the theorems are provided in the final subsections (Subsections~\ref{prf:pdf},  \ref{sec:cf_pf} and \ref{sec:prf_sphere_pack}) and can be omitted at a first reading.  We evaluate the  DF exponent on the Sato relay channel in Section~\ref{sec:num}. Finally, we conclude our discussion in Section~\ref{sec:concl} where we also mention several other   avenues of research. 
\section{Preliminaries}\label{sec:prelim}

\subsection{General Notation}
We generally adopt the notation from Csisz\'ar and K\"orner~\cite{Csi97} with a few minor modifications. Random variables  (e.g., $X$)   and their realizations (e.g., $x$)   are in capital and  small letters  respectively. All random variables take values on finite sets,  denoted in calligraphic font (e.g., $\calX$). For a sequence $x^n=(x_1,\ldots, x_n)\in\calX^n$, its {\em type} is the distribution $P(x)=\frac{1}{n}\sum_{i=1}^n\bone\{x=x_i\}$ where $\bone\{\mathrm{clause}\}$ is $1$ if the clause is true and $0$ otherwise.  All logs are with respect to base $2$ and we use the notation $\exp(t)$ to mean $2^t$.   Finally, $|a|^+ :=\max\{a, 0\}$ and $[a]:=\{1,\ldots, \ceil{ a}\}$ for any $a\in\bbR$. 

The set of distributions supported on $\calX$ is denoted as $\scP(\calX)$. The set of types in $\scP(\calX)$ with denominator $n$ is denoted as $\scP_n(\calX)$. The set of all sequences $x^n$ of type $P$ is the  {\em type class} of $P$ and is  denoted as $\calT_P :=\{x^n \in\calX^n : x^n \mbox{ has type } P\}$.  For a distribution $P\in\scP(\calX)$ and a stochastic matrix $V:\calX\to\calY$, we denote the joint distribution interchangeably as $P\cp V$ or $PV$. This should be clear from the context. For     $x^n \in\calT_P$,  the set of sequences $y^n \in\calY^n$ such that $(x^n, y^n)$ has joint type $P \cp V$ is the {\em $V$-shell} $\calT_V(x^n)$.  Let $\scV_n(\calY;P)$  be the family of stochastic matrices $V : \calX \to \calY$ for which the $V$-shell of a sequence of type $P\in\scP_n(\calX)$ is  not empty.  The elements of  $\scV_n(\calY;P)$   are called {\em conditional types compatible with $P$} (or simply conditional types if $P$ is clear from the context). 

Information-theoretic quantities are denoted in the usual way.  For example, $I(X;Y)$ and  $I(P,V)$ denote the mutual information where the latter expression makes clear that the joint  distribution of $(X,Y)$ is      $P\cp V$. In addition, $\hatI(x^n\wedge y^n)$ is the empirical mutual information of   $(x^n, y^n)$, i.e., if  $x^n   \in \calT_{P  }$ and $ y^n\in\calT_V(x^n)$, then,  $\hatI(x^n\wedge y^n)=I(P,V)$.  For a distribution $P\in\scP(\calX)$ and two stochastic matrices $V:\calX\to\calY$, $W:\calX\cp\calY\to\calZ$, $I( V, W|P)$ is the  conditional mutual information $I(Y;Z|X)$ where $(X,Y,Z)$ is distributed as  $P\cp V\cp W$.

We will also often use the asymptotic notation $\doteq $ to denote equality   to first-order in the exponent. That is, for two positive sequences $\{a_n, b_n\}_{n=1}^{\infty}$,  we  say that $a_n  \doteq b_n$ if and only if $ \lim_{n\to\infty}n^{-1}\log\frac{a_n}{b_n}= 0$. Also, we will use $\dotleq$ to denote inequality   to first-order in the exponent. That is, $a_n \dotleq b_n$ if and only if  $ \limsup_{n\to\infty}n^{-1}\log\frac{a_n}{b_n}\le 0$. Finally, $a_n = \Theta(b_n)$ if and only if there exists constants $0<c_1\le c_2 < \infty$ such that $c_1 b_n\le a_n \le c_2 b_n$ for $n$ sufficiently large. 

 \subsection{The Method of Types}
We also summarize some known facts about types that we use extensively in the sequel. The following lemma summarizes key results in \cite[Ch.~2]{Csi97}. 
\begin{lemma}[Basic Properties of Types] \label{lem:types}
Fix a type $P\in\scP_n(\calX)$ and sequence $x^n\in\scP_n(\calX)$. Also fix a conditional type $V\in\scV_n(\calY;P)$ and a sequence $y^n\in \calT_V(x^n)$.  For any stochastic matrix $W:\calX\to\calY$, we have 
\begin{enumerate}
\item  $|\scV_n(\calY;P) | \le (n+1)^{|\calX||\calY|}$ 
\item $(n+1)^{-|\calX||\calY|}\exp(n H(V|P)) \le |\calT_V(x^n)| \le \exp(n H(V|P))$
\item $W^n(y^n|x^n)  =\exp[-n (D(V \|W |P )+ H(V|P)) ]$
\item $(n+1)^{-|\calX||\calY|}\exp[-n  D(V \|W |P )]\le W^n(\calT_V(x^n)|x^n)  \le\exp[-n D(V \|W |P )] $
\end{enumerate}
\end{lemma}

The following lemmas are implicit in the results in \cite[Ch.~10]{Csi97}. We provide formal statements and proofs in Appendices~\ref{app:joint_typ} and \ref{app:joint_typ2}  for completeness.  Lemma~\ref{lem:joint_typ} is a conditional version of the following statement: Let $X^n$ be a length-$n$ sequence  drawn uniformly at random from $\calT_P$. The probability that an arbitrary sequence $\bary^n$ of marginal type $Q(y) =\sum_x P(x)V'(y|x)$  (marginal consistency) lies in the $V'$-shell of $X^n$ is roughly  $\exp[-n I(P,V')]$.  

\begin{lemma}[Joint Typicality  for Types I]\label{lem:joint_typ}
Let $P\in\scP_n(\calX_1)$, $V\in\scV_n(\calX_2;P)$ and $V'\in \scV_n(\calY;P\times V)$. Define $W(y|x_1):=\sum_{x_2} P(x_1) V(x_2|x_1)V'(y|x_1,x_2)$. Then for any $x_1^n\in\calT_P$, if $X_2^n$ is uniformly drawn from the shell $\calT_{ V}(x_1^n)$ and $\bary^n$ is any element of $\calT_W(x_1^n)$, 
\begin{equation}
\frac{1}{p_1(n)} \exp[-nI(V, V'|P)]\le \bbP\left[\bary^n \in \calT_{V'}(x_1^n, X_2^n) \right]\le p_2(n) \exp[-nI(V, V'|P)],
\end{equation}
where $p_1(n)$ and $p_2(n)$ are polynomial functions of $n$ depending only on the cardinalities of the alphabets.
\end{lemma}

Lemma~\ref{lem:joint_typ2} is a conditional version of the following statement: Let $\bary^n$ be a fixed length-$n$ sequence from $\calT_Q$. Let $X^n$ be a random length-$n$ sequence drawn uniformly at random from the marginal type class $\calT_P$ where  $P (x) = \sum_y Q(y) W(y|x)$ (marginal consistency). Then, the probability that $X^n$ lies in the $W$-shell of $\bary^n$ is   upper bounded by  $\exp[-nI(P,W)]$ up to a polynomial term.

\begin{lemma}[Joint Typicality  for Types II]\label{lem:joint_typ2}
Let  $P\in\scP_n(\calX_1)$, $V\in\scV_n(\calX_2;P)$ and $V'\in\scV_n(\calY;P)$. Let $W:\calY\times\calX_1\to\calX_2$ be any (marginally consistent) channel satisfying $\sum_y W(x_2|y,x_1) V' (y|x_1)= V(x_2|x_1)$. Fix $x_1^n \in \calT_P$. Let $X_2^n$ be uniformly distributed in $\calT_{V}(x_1^n)$. For any $\bary^n \in \calT_{V'}(x_1^n)$, we have
\begin{equation}
\bbP\left[X_2^n \in \calT_W(\bary^n, x_1^n)\right]\leq p_3(n)\exp [-n I(V', W|P)],
\end{equation}
where $p_3(n)$  is a  polynomial function of $n$ depending only on the cardinalities of the alphabets.
\end{lemma}


\subsection{The Relay Channel and Definition of Reliability Function}
In this section, we recall the definition of the relay channel and the notion of the reliability function for a channel.
\begin{definition}
A {\em $3$-node discrete memoryless relay channel (DM-RC)} is a tuple $(\calX_1\times\calX_2, W,\calY_2\times\calY_3)$ where $\calX_1,\calX_2,\calY_2$ and $\calY_3$ are finite sets and $W:\calX_1\times\calX_2\to\calY_2\times\calY_3$ is a stochastic matrix. The sender (node 1) wishes to communicate a message $M$ to the receiver (node 3) with the help of the relay node (node 2). See Fig.~\ref{fig:relay}.
\end{definition}
\begin{definition} \label{def:relay}
A $(2^{nR}, n)$-code for the DM-RC consists of a message set $\calM=[2^{nR}]$, an encoder $f: \calM\to\calX_1^n$ that assigns a codeword to each message, a sequence of relay encoders $g_i:\calY_2^{i-1}\to\calX_2,i\in [n]$ each  assigning a   symbol  to each past received sequence and a decoder $\varphi:\calY_3^n\to\calM$ that assigns an estimate of the message to each channel output.  The rate of this code is $R$. 
\end{definition} 
We denote the $i$-th component of the vector $f(m)\in\calX_1^n$ as $f_i(m)$.  We assume that $M$ is uniformly distributed on $\calM$ and the channel is memoryless. More precisely, this means that  the current received symbols $(Y_{2i}, Y_{3i})$ are conditionally independent of the message and the past symbols $(M, X_1^{i-1}, X_2^{i-1}, Y_2^{i-1}, Y_3^{i-1})$ given the current transmitted symbols $(X_{1i}, X_{2i})$, i.e., 
\begin{equation}
W^n(y_2^n, y_3^n | x_1^n, x_2^n) = \prod_{i=1}^n W(y_{2i}, y_{3i} | x_{1i}, x_{2i}).
\end{equation}
 
Fix a $(2^{nR}, n)$-code  given by the message set $\calM=[2^{nR}]$, encoder $f$, the relay encoders $(g_1, g_2,\ldots, g_n)$ and the decoder $\varphi$ yielding disjoint decoding regions $\calD_m =\varphi^{-1}(m) \subset\calY_3^n$. Let $g^n:\calY_2^n\to\calX_2^n$ denote the concatenation of the relay encoders, i.e., $g^n(y_2^n) := (x_2^*, g_2(y_{21}),g_3(y_2^2), \ldots, g_n(y_2^{n-1}))$, where $x_2^*\in\calX_2$ is any fixed symbol. (It does not matter which $x_2^* \in \calX_2$ is fixed because $n$ is allowed to tend to infinity.)   For a given  DM-RC  $W$ and  coding functions $(f, g^n, \varphi)$,  define $\bbP_W((Y_2^n,Y_3^n) = (y_2^n,y_3^n)|M=m)$ to be the probability that $(Y_2^n,Y_3^n) = (y_2^n,y_3^n)$ when message $m$ is sent under channel $W$, i.e., 
\begin{equation}
\bbP_W((Y_2^n,Y_3^n) = (y_2^n,y_3^n)|M=m):=\prod_{i=1}^n W(y_{2i}, y_{3i} | f_i(m), g_i(y_2^{i-1}) ). \label{eqn:def_PV}
 \end{equation} 
In addition, define the marginal probability 
\begin{equation}
\bbP_W(Y_3^n = y_3^n|M=m) := \sum_{y_2^n \in\calY_2^n}\bbP_W((Y_2^n,Y_3^n) = (y_2^n,y_3^n)|M=m). \label{eqn:margina}
\end{equation}

\begin{definition} \label{def:err_prb}
Let the {\em average error probability} given the code $(\calM, f,g^n,\varphi)$ be defined as 
\begin{equation}
\rmP_{\rme}(W; \calM, f, g^n, \varphi)  := \frac{1}{|\calM|}\sum_{m\in\calM}\bbP_W(Y_3^n \in\calD_m^c| M=m) , \label{eqn:def_err_prob}
\end{equation}
where $\calD_m^c :=\calY_3^n\setminus\calD_m$. We also denote the average error probability more succinctly as $\rmP_{\rme}(W)$ or $\bbP\big(\hatM\ne M \big)$ where the dependencies on the code are suppressed. 
\end{definition}


We now define the reliability function formally. 
\begin{definition} \label{def:er}
The reliability function~\cite{Csi97} for the DM-RC  $W$ is defined as 
\begin{equation}
 E(R):=\sup\,\, \left\{\liminf_{n\to\infty} \,\, \frac{1}{n}\log \frac{1}{ \bbP\big(\hatM\ne M\big)  } \right\} \label{eqn:tilE}
\end{equation}
where $M\in\calM:=[2^{nR}]$ and the supremum is   over all sequences of $(2^{nR},n)$  codes for the DM-RC.
\end{definition}

%


As in~\cite{Bra12}, we use block-Markov coding to send a message $M$ representing $N \Reff=nb\Reff$ bits of information over the DM-RC. We use the channel $N$ times and this total blocklength  is partitioned into $b$ correlated blocks each of  length $n$. We term $n$, a large integer, as the {\em per-block blocklength}. The number of blocks $b$ is fixed and regarded as a constant (does not grow with $n$). We discuss  the effect of $b=b_n$ growing in Section~\ref{sec:rmk_pdf}. The message is split into $b-1$ sub-messages $M_j, j \in [b-1]$, each representing $nR$  bits of information. Thus, the {\em effective rate} of the code is  
\begin{equation}
\Reff = \frac{(b-1)nR}{N} = \frac{(b-1)nR}{nb} = \frac{b-1}{b}R.   \label{eqn:reff}
\end{equation}
We also say that $R>\Reff$ is the {\em per-block rate}.  Under this  coding setup, we wish to provide  lower  bounds on  the reliability function.  We also prove an upper bound on the reliability function.

\subsection{Background on Error Exponents via the Method of Types} \label{sec:review}
In this section, we provide a brief summary of Csisz\'ar-K\"orner-style~\cite{Csi97} error exponents for   channel coding~\cite{CKM} and lossy source coding~\cite{Marton74}. For a more comprehensive exposition, see Csisz\'ar's review paper~\cite{Csis00}.

For channel coding, the packing lemma~\cite[Lem.~10.1]{Csi97} allows us to show that  for every distribution $P \in\scP(\calX)$, the following exponent, called the {\em random coding error exponent}, is achievable
\begin{align}
E_{\rmr}(R, P ) := \min_{V:\calX\to\calY} D(V \| W|P ) + |I(P,V)-R|^+. \label{eqn:rcee}
\end{align}
Roughly speaking, the term $D(V\| W|P)$ represents the atypicality of the channel $W:\calX\to\calY$, namely that it behaves like $V$. (More precisely, for finite $n$, the {\em conditional type} of $y^n$ given $x^n$ is  represented by $V$.)  The term $|I(P,V)-R|^+$ represents the deviation of the code rate from the rate that the channel can support  $I(P,V)$. Besides the proof using the packing lemma~\cite[Thm.~10.2]{Csi97}, one can show that $E_{\rmr}(R,P)$ is achievable by Gallager's Chernoff bounding techniques~\cite[Ch.~5]{gallagerIT} or by considering a family of decoding rules which involve maximizing a function $\alpha(P,V )$ \cite{CK81}. Here, $P$ represents the   type of the codewords of a constant composition code and $V$ represents the conditional type of the channel output given  a particular codeword. If $\alpha(P,V)=I(P,V)$ this corresponds to MMI decoding; if $\alpha(P,V)=H(V|P)$, this corresponds to minimum conditional entropy decoding (these two decoding strategies are identical for constant composition codes); while if $\alpha(P,V) = D(V\| W|P)+ H(V|P)$ this corresponds to ML decoding. Notice that ML decoding depends on knowledge of the true channel $W$.  For PDF, we will use MMI decoding and obtain an exponent that is similar to \eqref{eqn:rcee}. For CF, we find it convenient to use a combination of MMI and ML decoding in addition to other techniques such as the one-at-a-time union bound described in Section~\ref{sec:oneat}.

It was shown by Haroutunian~\cite{Haroutunian68} that for every distribution $P\in\scP(\calX)$, the following is an upper bound to the reliability function
\begin{equation}
E_{\mathrm{sp}}(R,P) := \min_{V:\calX\to\calY:  I(P,V)\le R} D(V\| W|P). \label{eqn:sphere}
\end{equation}
This is customarily called the {\em sphere-packing exponent}. An alternative expression is given by Shannon-Gallager-Berlekamp~\cite{sgb}.  In the presence of feedback, Haroutunian~\cite{Har77} also proved the following upper bound to the reliability function
\begin{equation}
E_{\rmH}(R) := \min_{V:\calX\to\calY  : \rvC(V) \le R}\max_P D(V\| W|P) \label{eqn:har}
\end{equation}
where $\rvC(V):=\max_P I(P,V)$ is the Shannon capacity of $V$.  This is   called the {\em Haroutunian exponent}.  It is known that $E_{\rmH}(R) = \max_P E_{\mathrm{sp}}(R,P)$ for output symmetric channels but in general  $E_{\rmH}(R)> \max_P E_{\mathrm{sp}}(R,P)$ \cite{Har77}. Our upper bound on the reliability function is similar to the Haroutunian exponent with the exception that $\rvC(V)$ in \eqref{eqn:har} is replaced by the cutset bound $\rvC_{\mathrm{cs}}(V): = \max_{P_{X_1 X_2}} \min\{ I(X_1 X_2; Y_3 ), I(X_1; Y_2 Y_3| X_2) \}$ where $V$ is the conditional distribution of $(Y_2, Y_3)$ given $(X_1, X_2)$. 

In our proof of the CF exponent in Section~\ref{sec:cf}, we will make use of a technique Marton~\cite{Marton74} developed to analyze the error exponent for compressing discrete memoryless  sources with a fidelity criterion. In contrast to channel coding, this exponent is known for all rates and is given by  the {\em Marton exponent}
\begin{equation}
F(P, R, \Delta) = \inf_{Q : R(Q,\Delta) > R} D(Q\| P) \label{eqn:marton_exp}
\end{equation}
where $P\in\scP(\calX)$ is the distribution of the source, $R$ is code rate, $\Delta$  is the distortion level and $R(Q,\Delta)$ is the rate-distortion function.  Marton's exponent in~\eqref{eqn:marton_exp} is intuitive in view of Sanov's theorem~\cite[Ch.~2]{Csi97}: If the coding rate $R$ is below    $R(Q,\Delta)$ for  some $Q$, which represents the type of the realization of the source $X^n\sim P^n$, then the error decays with exponent $D(Q\| P)$.  A key lemma used to prove the direct part of \eqref{eqn:marton_exp} is called the {\em type covering lemma} proved by Berger~\cite{berger71}. Also see~\cite[Lem.~9.1]{Csi97}. We use a refined version of this technique and combine that with other packing techniques to prove the CF exponent in Section~\ref{sec:cf}.

\subsection{The ``One-At-A-Time Union Bound''} \label{sec:oneat} 

As mentioned in the Introduction, we use a modification of the union bound in our proof of the error exponent for CF. This is called (in our words) the {\em one-at-a-time union bound} and was used for error exponent analyses in~\cite{Sca12, Sca13Nov}. To describe this technique, first recall  the {\em truncated union bound} which says that if $\calE_m, 1\le m \le M$ is a collection of events with the same probability, then 
\begin{equation} \label{eqn:truc}
\bbP\bigg(\bigcup_{m=1}^M \calE_m \bigg) \le \min\big\{ 1, M\bbP(\calE_1)\big\}.
\end{equation}
The one-at-a-time union bound  says that if we have two independent sequences of identically distributed   random variables $A_m, 1\le m \le M$ and $B_k , 1\le k \le K$, 
then the probability that any of the pairs $(A_m ,  B_k)$ belongs to some set $\calD$ can be bounded as 
\begin{align}
&\bbP\bigg(\bigcup_{m=1}^{M}\bigcup_{k=1}^K \big\{(A_m, B_k) \in\calD\big\}\bigg)\nn\\*
& \le \min\bigg\{ 1, M \bbP\bigg( \bigcup_{k=1}^K \big\{ (A_1 , B_k) \in\calD\big\}\bigg) \bigg\} \label{eqn:use_trunc} \\
& = \min\bigg\{ 1, M  \bbE\bigg[\bbP\bigg( \bigcup_{k=1}^K \big\{ (A_1 , B_k) \in\calD \big\}  \bigg|A_1\bigg)\bigg] \bigg\}\\
& \le \min\Big\{ 1, M  \bbE\Big[\min\{ 1, K  \bbP\big(   \big\{(A_1 , B_1) \in\calD\big\} \big|A_1 \big)\Big] \Big\} \Big\}.\label{eqn:one_time}
\end{align}
In \eqref{eqn:use_trunc} and \eqref{eqn:one_time}, we applied the truncated union bound in \eqref{eqn:truc}. Clearly, applying the union bounds in the other order ($k$ first, then $m$) yields  another upper bound. These bounds are usually better than a simple application of the union bound on both indices jointly, i.e.,
\begin{equation}
\bbP\bigg( \bigcup_{m,k} \big\{(A_m, B_k) \in\calD\big\}\bigg)\le \min\{1, MK\bbP(\big\{(A_1, B_1) \in\calD\big\})\}.
\end{equation}

\section{Partial Decode-Forward (PDF)}\label{sec:dec}

We warm up by deriving two achievable error exponents using PDF. In  PDF, the relay decodes part of the message in each block. For block $j\in [b]$, the part of the message that is decoded by the relay is indicated as $M_j'$ and the remainder of the message  is $M_j''$. Thus, $M_j = (M_j', M_j'')$.  We will state the main theorem  in Section~\ref{sec:pdf}, provide some remarks in Section~\ref{sec:rmk_pdf}   and prove it in Section~\ref{prf:pdf}.

\subsection{Analogue of the Random Coding Error Exponent} \label{sec:pdf}

The analogue of the random coding exponent is presented as follows:
\begin{theorem}[Random Coding Error Exponent for Partial Decode-Forward]  \label{thm:pdf}
Fix $b\in\bbN$,  auxiliary alphabet $\calU$ and a  rate $R$. Let $R = R'+R''$ for two non-negative rates $R'$ and $R''$. Fix a joint  distribution $Q_{X_2}\times  Q_{U|X_2} \times   Q_{X_1|U X_2}\in\scP(\calX_2\times\calU\times\calX_1)$.
These distributions induce the following virtual channels:
\begin{align}
W_{Y_2|U X_2}(y_2|u,x_2) &:=\sum_{x_1,y_3} W(y_2,y_3|x_1,x_2)Q_{X_1|U X_2}(x_1|u, x_2) ,\label{eqn:vir1}\\
W_{Y_3|U X_2}(y_3|u,x_2) &:=  \sum_{x_1,y_2} W(y_2,y_3|x_1,x_2)Q_{X_1|U X_2}(x_1|u, x_2), \label{eqn:vir2}\\
W_{Y_3|U X_1 X_2} (y_3|u,x_1,x_2) & :=\sum_{y_2} W(y_2,y_3|x_1,x_2) ,\qquad\forall \, u\in\calU. \label{eqn:vir3}
\end{align}
The following is a lower bound on the reliability function: 
\begin{eqnarray}
E(\Reff) \ge E_{\mathrm{pdf}}^{(b)}(\Reff)  := \frac{1}{b} \min\{F(R'), G (R'),  \tilG(R'')\} ,  \label{eqn:pdf_lb}
\end{eqnarray}
where $F(R'), G(R' )$ and  $\tilG(R'')$ are  constituent error exponents defined as 
\begin{align}
F(R')&:=\min_{V:\calU\times\calX_2\to\calY_2}D(V\|W_{Y_2|U X_2}|Q_{U X_2})  + \left|  I(Q_{U|X_2},V|Q_{X_2})-R' \right|^+  \label{eqn:F}\\
G(R')&:=\min_{V:\calU\times\calX_2\to\calY_3}D(V\|W_{Y_3|U X_2}|Q_{U X_2})  +\left|    I(Q_{ U , X_2},V)-R'  \right|^+ \label{eqn:G1}\\
 \tilG(R'')&:= \min_{V:\calU\times\calX_1\times\calX_2\to\calY_3}  D(V\|W_{Y_3|U X_1 X_2}|Q_{U X_2 X_1} ) +\left|   I(Q_{X_1|U X_2},V|Q_{U X_2})-R''  \right|^+  \label{eqn:G2}
\end{align}
\end{theorem}
The proof of this result is based on a modification of the techniques used in the packing lemma~\cite{Csi97, Csis00, Kor80b, CKM, Har08} and is provided in Section~\ref{prf:pdf}. 

\subsection{Remarks on the Error Exponent for  Partial Decode-Forward } \label{sec:rmk_pdf}
A few comments are in order concerning Theorem~\ref{thm:pdf}.

\begin{enumerate}

\item  \label{item:Rp} Firstly, since $Q_{X_2}$, $Q_{U|X_2}$ and $Q_{X_1| U X_2}$ as well as the splitting of $R$ into $R'$ and $R''$  are arbitrary, we can maximize the lower bounds in \eqref{eqn:pdf_lb}   over these free parameters.   Particularly, for a fixed split $R'+R''=R$, if 
\begin{align}
R' &<  I(U  ;Y_2|X_2) \\
R' &< I(U X_2;Y_3 )\\
R'' &< I( X_1;Y_3|U X_2),
\end{align}
for some $Q_{X_2}$, $Q_{U|X_2}$ and $Q_{X_1| U X_2}$, then $F(R')$, $G(R')$ and $\tilG(R'')$ are positive.   Hence, the error probability decays exponentially fast if $R$ satisfies the PDF lower bound in~\eqref{eqn:cap_pdf}.  In fact, $|\calU|$ is also a free parameter. As such we may let $|\calU|\to \infty$. It is not obvious that a finite $\calU$ is optimal (in the sense that $E_{\mathrm{pdf}}^{(b)}(\Reff)$ does not  improve by increasing $|\calU|$). This is because   we cannot apply the usual cardinality bounding techniques based on the support lemma~\cite[App.~C]{elgamal}. A method to prove that the cardinalities of auxiliary random variables can be bounded for error exponents was presented in \cite[Thm.~2]{Liu96} but the technique is specific to the multiple-access channel and  does not appear to   carry over  to our setting in a straightforward manner.  

\item  \label{item:df} Secondly, we can interpret $F(R')$ as the error exponent  for decoding a part of the message $M_j'$ (block $j$) at the relay and $G(R')$  and $\tilG(R'')$  as the error exponents for decoding the whole of the message $M_j = (M_j', M_j'')$ at the receiver.  Setting $U=X_1$ and $R''=0$ recovers DF for which the error exponent (without the sliding-window modification) is provided in~\cite{Bra12}. Indeed, now we only have two exponents $F(R)$ and $G(R)$ corresponding respectively to the errors at the relay and the decoder. For the overall exponent to be positive, we require 
\begin{equation}
R< \min\{ I(X_1;Y_2|X_2),I(X_1X_2; Y_3)\}
\end{equation}
which corresponds exactly to the DF lower bound~\cite[Thm.~16.2]{elgamal}. However, the form of the exponents is different from that in \cite{Bra12}. Ours is in the Csisz\'ar-K\"orner~\cite{Csi97} style while \cite{Bra12} presents exponents in the Gallager~\cite{gallagerIT} form. Also see item \ref{item:gallager} below.

\item   \label{item:b} The exponent in~\eqref{eqn:pdf_lb}   demonstrates a tradeoff between the effective rate and error probability: for a fixed $R$, as the number of blocks $b$ increases, $\Reff$ increases but because of the division by $b$, $E_{\mathrm{pdf}}^{(b)}(\Reff)$ decreases.  Varying $R$ alone, of course, also allows us to observe  this tradeoff.  Now in capacity analysis, one usually takes the number of blocks $b$ to tend to infinity so as to be arbitrary close to a $*$-forward lower bound \cite[Sec.~16.4.1]{elgamal}. If we let $b$ increase with  the per-block blocklength $n$, then the decay in the error probability would no longer be exponential  because  we divide by $b$ in \eqref{eqn:pdf_lb}  so $E_{\mathrm{pdf}}^{(b)}(\Reff)$ is arbitrarily small for large enough $n$. For example if $b_n = \Theta(n^{1/2})$, then the error probability decays as $\exp(-\Theta(n b_n^{-1}))=\exp(-\Theta(n^{1/2}))$. However, the effective rate $\Reff$ would come arbitrarily close to the per-block rate $R$. If $R$ is also allowed to tend towards a  $*$-forward lower bound, this would be likened to operating in the moderate-    instead of large-deviations regime as the rate of decay of $\rmP_{\rme}(W)$ is subexponential but the effective rate of the code is arbitrarily close to a  $*$-forward lower bound. See~\cite{Tan12,altug10} for a partial list of works on moderate-deviations in information theory.

\item    \label{item:sliding} In general, the sliding window technique \cite{Carleial,KGG05} may yield  potentially better exponents. We do not explore this extension here but note that the improvements can be obtained by appealing to the techniques in~\cite[Props.~1 and~2]{Bra12}. In the sliding window decoding technique, the receiver estimates a message such that two typicality conditions are simultaneously satisfied. See \cite[Sec.~18.2.1, pp.~463]{elgamal}. This corresponds, in our setting, to maximizing two separate empirical mutual information quantities simultaneously, which is difficult to analyze. 

\item \label{item:expu} We also derived expurgated error exponents   for PDF using the technique outlined in \cite[Ex.~10.2 and 10.18]{Csi97} together with  ML decoding. These appear similarly to their classical forms and so are not included in this paper. 
 
\item  \label{item:random} In the proof, we use a random coding argument and show that averaged over a random code ensemble, the probability of error is desirably small. We do not first assert the existence of a good codebook via the classical packing lemma~\cite[Lem.\ 10.1]{Csi97} (or its variants~\cite{Csis00, Kor80b, CKM, Har08}) then upper bound the  error probability {\em given the non-random codebook}.   We also use a change-of-measure technique that was also used  by Hayashi~\cite{Hayashi09} for proving second-order coding rates in channel coding.  This change-of-measure technique allows us to use a random  constant composition code ensemble  and subsequently analyze this ensemble as if it were an i.i.d.\ ensemble, thus simplifying the analysis. 

\item \label{item:gallager} As is well known~\cite[Ex.~10.24]{Csi97}, we may lower bound the exponents in \eqref{eqn:F}--\eqref{eqn:G2} in the Gallager form~\cite{gallagerIT, gallager_MAC}, which is more amenable to computation. Furthermore, the Gallager-form  lower bound is tight for capacity-achieving input distributions in the point-to-point setting. For the DM-RC using the PDF scheme,   for a fixed  joint  distribution $Q_{X_2}\times  Q_{U|X_2} \times   Q_{X_1|U X_2}\in\scP(\calX_2\times\calU\times\calX_1)$,
\begin{align}
F(R')  &\ge \max_{\rho \in [0,1]} \left\{ -\rho R' -  \log \sum_{x_2, y_2} Q_{X_2}(x_2) \bigg[ \sum_u Q_{U|X_2}(u|x_2)W_{Y_2| U X_2}(y_2|u,x_2)^{\frac{1}{1+\rho}}\bigg]^{1+\rho}\right\},  \label{eqn:gal1}\\
G(R') &\ge \max_{\rho \in [0,1]} \left\{ -\rho R' -  \log \sum_{ y_3} \bigg[ \sum_{u,x_2} Q_{U X_2}(u,x_2)W_{Y_3| U X_2}(y_3|u,x_2)^{\frac{1}{1+\rho}}\bigg]^{1+\rho}\right\}, \label{eqn:gal2} \\
\tilG(R'')& \ge \max_{\rho \in [0,1]} \left\{ -\rho R'' -  \log \sum_{ u,x_2,y_3}Q_{UX_2}(u, x_2) \bigg[ \sum_{ x_1   } Q_{ X_1|U X_2}( x_1|u,x_2)W_{Y_3| U X_1  X_2}(y_3|u,x_1, x_2)^{\frac{1}{1+\rho}}\bigg]^{1+\rho}\right\} . \label{eqn:gal3}
\end{align}
We compute these exponents for the Sato relay channel~\cite{sato76} in Section~\ref{sec:num}.
\end{enumerate}

\subsection{Proof of  Theorem~\ref{thm:pdf}} \label{prf:pdf}

\begin{proof}
We code over $b$ blocks each of length $n$ (block-Markov coding~\cite{CEG}). Fix rates $R'$ and $R''$ satisfying $R'+R''=R$.  We fix a joint type $Q_{X_2}\times  Q_{U|X_2} \times   Q_{X_1|U X_2}  \in \scP_n(\calX_2\times\calU\times\calX_1)$. 
Split   each message $M_j , j\in[b-1]$ of rate $R$ into two independent parts  $M_j'$ and $M_j''$ of rates $R'$ and $R''$ respectively.

Generate $k' =\exp(nR')$ sequences $x_2^n(m_{j-1}'), m_{j-1}' \in [k'] $ uniformly at random from the type class $\calT_{Q_{X_2}}$. For each $m_{j-1}' \in [k']$ (with $m_0'=1$), generate $k'$ sequences $u^n(m_j' | m_{j-1}')$ uniformly at random from the $Q_{U|X_2}$-shell $\calT_{Q_{U|X_2}} ( 	x_2^n(m_{j-1}'))$.  Now for every $(m_{j-1}', m_j') \in [k']^2$, generate $k'' = \exp(nR'')$ sequences $x_1^n(m_j', m_j''|m_{j-1}')$ uniformly at random from the $Q_{X_1|U X_2}$-shell  $\calT_{Q_{X_1|U X_2} } ( u^n(m_j' | m_{j-1}'), 	x_2^n(m_{j-1}') ) $. This gives a random codebook. Note that the $x_2^n(m_{j-1}')$ sequences need not be distinct if $R' \ge H(Q_{X_2})$ (and similarly for the  $u^n(m_j' | m_{j-1}')$  and $x_1^n(m_j', m_j''|m_{j-1}')$ sequences)  but this does not affect the subsequent arguments.  We bound the error probability averaged  over realizations of this random codebook.

The sender and relay cooperate to send $m_j'$ to the receiver. In block $j$, relay transmits $x^n(m_{j-1}')$ and   transmits $x_1^n(m_j', m_j''|m_{j-1}')$ (with $m_0'=m_b'=1$ by convention).

At the $j$-th step, the relay does MMI decoding~\cite{Goppa} for $m_j'$ given $m_{j-1}'$ and $y_2^n(j)$. More precisely, it declares that $\check{m}_j'$ is sent if 
\begin{align}
\check{m}_j' = \argmax_{m_j' \in [ \exp(n (R' ) ) ] }  \hatI( u^n (m_j'|m_{j-1}') \wedge y_2^n(j) | x_2^n(m_{j-1}') ) . \label{eqn:relay_j}
\end{align}
By convention, set $m_0'=1$. Recall that $\hatI( u^n (m_j'|m_{j-1}') \wedge y_2^n(j) | x_2^n(m_{j-1}') )$ is the conditional mutual information $I(\tilU; \tilY_2| \tilX_2)$ where the dummy random variables $(\tilX_2, \tilU, \tilY_2)$ have joint type given by the sequences $(u^n (m_j'|m_{j-1}'), x_2^n(m_{j-1}'), y_2^n(j))$. 
After all blocks are received, the decoder performs backward decoding~\cite{Zeng89, Willems85}  by using the MMI decoder~\cite{Goppa}.  In particular, it declares that $\hatm_{j}'$ is sent if 
\begin{equation}
\hatm_{j}' = \argmax_{m_j' \in [ \exp(n (R' ) ) ]   }    \hatI( u^n (m_{j+1}'|m_{j}') , x_2^n(m_j')\wedge y_3^n(j+1) ) . \label{eqn:decoder_j}
\end{equation}
After all $\hatm_j', j\in [b-1]$ have been decoded in step~\eqref{eqn:decoder_j},  the decoder then decodes   $m_j''$    using MMI~\cite{Goppa}  as follows:
\begin{align}
\hatm_j''=\argmax_{m_j''\in[ \exp(n (R'' ) ) ]  }   \hatI(   x_1^n(m_{j}', m_j''|m_{j-1}') \wedge y_3^n(j)     | \, u^n (m_{j}'|\hatm_{j-1}') , x_2^n(\hatm_{j-1}') ) \label{eqn:decoder_j2} .
 \end{align}
In steps~\eqref{eqn:relay_j}, \eqref{eqn:decoder_j} and \eqref{eqn:decoder_j2} if there exists more than one message attaining the $\argmax$, then pick any one uniformly at random.   We assume, by symmetry, that $M_j=(M_j',M_j'')=(1,1)$ is sent for all $j\in[b - 1]$. The line of analysis in \cite[Sec.\ III]{Bra12} yields  
\begin{equation} \label{eqn:union_bd}
\bbP(\hatM\ne M)\le (b-1) (\epsilon_{\rmR}+\epsilon_{\rmD,1} +\epsilon_{\rmD,2}) 
\end{equation}
where  for any $j\in [b-1]$,
\begin{equation}
\epsilon_{\rmR}: =\bbP( \check{M}_j'\ne 1|\check{M}_{j-1}'= 1)
\end{equation}
 is the error probability  in decoding  at the relay and for any $j\in [b-1]$,
\begin{align}
 \epsilon_{\rmD,1} &: = \bbP( \hatM_{j+1}'\ne 1| \hatM_{j+1}'=1, \check{M}_{j+1}=1), \quad \mbox{and}\\
    \epsilon_{\rmD,2}&:=\bbP( \hatM_{j}''\ne 1| \hatM_{j}'=1, \check{M}_{j-1}'=1)
\end{align} 
and are the error probabilities of decoding $M_{j+1}'$ and $M_{j}''$ at the decoder. Since $b$ is assumed to be constant, it does not affect the exponential dependence of the error probability in \eqref{eqn:union_bd}. So we just bound $\epsilon_{\rmR}$, $\epsilon_{\rmD,1}$ and $\epsilon_{\rmD,2}$. Since all the calculations are similar, we   focus on
$\epsilon_\rmR$ leading to the error exponent $F(R')$ in \eqref{eqn:F}.

An error occurs in step~\eqref{eqn:relay_j} (conditioned on neighboring blocks being decoded correctly so $\check{m}_{j-1}'={m}_{j-1}'=1$) if and only if there exists some index $\tilm_j'\ne 1$ such that the empirical conditional information computed with respect to $u^n(\tilm_j'|1)$ is higher than that of $u^n(1|1)$, i.e., 
\begin{equation}
\epsilon_\rmR=\bbP( \exists\, \tilm_j'\ne 1: \hatI(U^n(\tilm_j'|1)\wedge Y_2^n(j) | X_2^n(1))\ge \hatI(U^n(1|1)\wedge Y_2^n(j) | X_2^n(1))) . 
\end{equation}
We can condition this on various values of $( u^n,x_2^n)\in\calT_{Q_{UX_2 }}$ as follows,
\begin{equation}
\epsilon_\rmR=\sum_{(u^n,x_2^n) \in \calT_{Q_{U X_2}}}\frac{1}{| \calT_{Q_{U X_2}}|}\beta_n(u^n,x_2^n),
\end{equation}
where 
\begin{align}
\beta_n(u^n,x_2^n):=\bbP \big( \exists\, \tilm_j'\ne 1:\,  &\hatI(U^n(\tilm_j'|1)\wedge Y_2^n(j) | X_2^n(1)) \nn\\
&\ge \hatI(U^n(1|1)\wedge Y_2^n(j) | X_2^n(1)) \, \big|\, (U^n(1|1),X_2^n(1))= (u^n, x_2^n) \big).
\end{align}
It can be seen that $\beta_n(u^n,x_2^n)$ does not depend on $(u^n, x_2^n )\in \calT_{Q_{UX_2 }}$ so we simply write $\beta_n = \beta_n(u^n, x_2^n)$ and we only have to upper bound $\beta_n$. Because $x_1^n(1,1|1)$ is drawn uniformly at random from $\calT_{Q_{X_1|UX_2}}(u^n,x_2^n)$,
\begin{align}
\beta_n= \sum_{x_1^n \in \calT_{Q_{X_1|UX_2}}(u^n,x_2^n)}  \frac{1}{|\calT_{Q_{X_1|UX_2}}(u^n,x_2^n)|}  \sum_{y_2^n}W^n(y_2^n|x_1^n, x_2^n)\mu_n(y_2^n).
\end{align}
where 
\begin{align}
\mu_n(y_2^n):=\bbP \big( \exists\, \tilm_j'\ne 1:\,  &\hatI(U^n(\tilm_j'|1)\wedge Y_2^n(j) | X_2^n(1)) \nn\\
&\ge \hatI(U^n(1|1)\wedge Y_2^n(j) | X_2^n(1)) \, \big|\, (U^n(1|1),X_2^n(1))= (u^n, x_2^n), Y_2^n(j)=y_2^n \big). \label{eqn:mu_n}
\end{align}
Note that the event before the conditioning in $\mu_n(y_2^n)$ does not depend on  $\{X_1^n(1 , 1|1)=x_1^n\}$ so we drop the dependence on $x_1^n$  from the notation  $\mu_n(y_2^n)$.  Intuitively  $(U^n(1|1), X_2^n(1)) = (u^n,x_2^n)$ is the ``cloud center''  and $X_1^n(1,1|1)=x_1^n$ the ``satellite codeword''. Since $x_1^n \in  \calT_{Q_{X_1|UX_2}}(u^n,x_2^n)$, our knowledge of the precise $x_1^n$ does not increase our knowledge of $U^n(\tilm_j' | 1), \tilm_j'\ne 1$, which is the only source of randomness in the probability in~\eqref{eqn:mu_n}. 
We continue to bound $\beta_n$ as follows:
\begin{align}
\beta_n &\le \sum_{x_1^n \in \calT_{Q_{X_1|UX_2}}(u^n,x_2^n)}(n+1)^{|\calU||\calX_1||\calX_2|} \exp(-nH(Q_{X_1|UX_2}| Q_{UX_2}))\sum_{y_2^n}W^n(y_2^n|x_1^n, x_2^n)\mu_n(y_2^n) \label{eqn:lower_bd_tc3}\\
&= \sum_{x_1^n \in \calT_{Q_{X_1|UX_2}}(u^n,x_2^n)}(n+1)^{|\calU||\calX_1||\calX_2|} Q_{X_1|UX_2}^n(x_1^n|u^n,x_2^n) \sum_{y_2^n}W^n(y_2^n|x_1^n, x_2^n)\mu_n(y_2^n) \label{eqn:prob_TC} \\
&\le (n+1)^{|\calU||\calX_1||\calX_2|}  \sum_{y_2^n}\mu_n(y_2^n)\sum_{x_1^n  }Q_{X_1|UX_2}^n(x_1^n|u^n,x_2^n) W^n(y_2^n|x_1^n, x_2^n) \label{eqn:drop_terms}\\
&\le(n+1)^{|\calU||\calX_1||\calX_2|}   \sum_{y_2^n}   W_{Y_2|UX_2}^n(y_2^n|u^n, x_2^n) \mu_n(y_2^n) \label{eqn:bound_beta} ,
\end{align}
where \eqref{eqn:lower_bd_tc3} follows by lower bounding the size of  $\calT_{Q_{X_1|UX_2}}(u^n,x_2^n)$ (Lemma~\ref{lem:types}), \eqref{eqn:prob_TC} follows by the fact that the $Q_{X_1|UX_2}^n(\fndot|u^n,x_2^n)$-probability of any sequence in $\calT_{Q_{X_1|UX_2}}(u^n,x_2^n)$ is exactly $\exp(-nH(Q_{X_1|UX_2}| Q_{UX_2}))$ (Lemma~\ref{lem:types}),  \eqref{eqn:drop_terms} follows by dropping the constraint $x_1^n \in \calT_{Q_{X_1|UX_2}}$, and \eqref{eqn:bound_beta} follows from the definition of $W_{Y_2|UX_2}$ in \eqref{eqn:vir1}.  We have to perform the calculation  leading to \eqref{eqn:bound_beta} because each $X_1^n$ is drawn uniformly at random from $\calT_{Q_{X_1|UX_2}}(u^n,x_2^n)$ and not from the product measure $Q_{X_1|UX_2}^n(\fndot|u^n, x_2^n)$, which would simplify the calculation and the introduction of the product  (memoryless) channel $W_{Y_2|UX_2}^n$. This change-of-measure technique (from constant-composition to product) was also used by Hayashi~\cite[Eqn.~(76)]{Hayashi09} in his work on second-order coding rates for channel coding. 

Hence it essentially remains   to bound $\mu_n(y_2^n)$ in \eqref{eqn:mu_n}. By applying the union bound, we have 
\begin{equation}
\mu_n(y_2^n)\le\min\{1,\exp(nR')\tau_n(y_2^n)\} \label{eqn:mu_def}
\end{equation}
where 
\begin{align}
\tau_n(y_2^n):=\bbP \big(    & \hatI(U^n(2|1)\wedge Y_2^n(j) | X_2^n(1)) \nn\\
&\ge \hatI(U^n(1|1)\wedge Y_2^n(j) | X_2^n(1)) \, \big|\, (U^n(1|1),X_2^n(1))= (u^n, x_2^n), Y_2^n(j)=y_2^n \big).
\end{align}
The only randomness now is in  $U^n(2|1)$. Denote the conditional type of $y_2^n$ given $x_2^n$ as $P_{y_2^n|x_2^n}$. Now consider reverse channels $\tilV:\calY_2\times\calX_2\to\calU$ such that  $\sum_{y_2} \tilV(u |y_2,x_2)P_{y_2^n|x_2^n}(y_2|x_2)=Q_{U|X_2}(u|x_2)$ for all $(u,x_2, y_2)$, i.e., they are marginally consistent with $Q_{U|X_2}$. Denote this class of reverse channels as $\scR(Q_{U|X_2})$.    We have,
\begin{align}
\tau_n(y_2^n)=\sum_{\substack{\tilV \in \scV_n (\calU ; P_{y_2^n|x_2^n} Q_{X_2} )  \cap\scR(Q_{U|X_2}) \\ I( P_{y_2^n|x_2^n} ,\tilV | Q_{X_2} ) \ge \hatI( u^n\wedge y_2^n |x_2^n) }} \bbP( U^n(2|1) \in \calT_{\tilV}( y_2^n, x_2^n) ).
\end{align}
By Lemma~\ref{lem:joint_typ2} (with the identifications $P\leftarrow Q_{X_2}$, $V\leftarrow Q_{U|X_2}$, $V'\leftarrow P_{y_2^n|x_2^n}$ and $W\leftarrow \tilV$),  for any $\tilV\in\scR(Q_{U|X_2})$,
\begin{align}
\bbP( U^n(2|1) \in \calT_{\tilV}( y_2^n, x_2^n) ) \dotleq\exp(-n  I( P_{y_2^n|x_2^n} ,\tilV | Q_{X_2} ) ).
\end{align}
 As a result,
\begin{align}
\tau_n(y_2^n)  &\dotleq\sum_{\substack{\tilV \in \scV_n (\calU ; P_{y_2^n|x_2^n} Q_{X_2} )  \cap\scR(Q_{U|X_2}) \\ I( P_{y_2^n|x_2^n} ,\tilV | Q_{X_2} ) \ge \hatI( u^n\wedge y_2^n |x_2^n) }} \exp(-n  I( P_{y_2^n|x_2^n} ,\tilV | Q_{X_2} ) )\\
&\dotleq\sum_{\substack{\tilV \in \scV_n (\calU ; P_{y_2^n|x_2^n} Q_{X_2} )  \cap\scR(Q_{U|X_2})  \\ I( P_{y_2^n|x_2^n} ,\tilV | Q_{X_2} ) \ge \hatI( u^n\wedge y_2^n |x_2^n) }} \exp(-n   \hatI( u^n\wedge y_2^n |x_2^n))\\
&\doteq \exp(-n   \hatI( u^n\wedge y_2^n |x_2^n))
\end{align}
Plugging this back into~\eqref{eqn:mu_def} yields,
\begin{align}
\mu_n(y_2^n) &\dotleq \min\left\{1,  \exp( -n (   \hatI( u^n\wedge y_2^n |x_2^n)-R'))   \right\} \\*
 &\doteq\exp\left[ -n  \left|   \hatI( u^n\wedge y_2^n |x_2^n)-R' \right|^+   \right]  .
\end{align}
Plugging this back into \eqref{eqn:bound_beta} yields,
\begin{align}
\beta_n &\dotleq \sum_{y_2^n} W_{Y_2|UX_2}^n(y_2^n|u^n, x_2^n) \exp\left[ -n  \left|   \hatI( u^n\wedge y_2^n |x_2^n)-R' \right|^+   \right]  \\
 &= \sum_{V\in\scV_n(\calY_2;Q_{UX_2})} W_{Y_2|UX_2}^n(\calT_V(u^n,x_2^n) |u^n, x_2^n) \exp\left[ -n  \left|   I( Q_{U|X_2},V|Q_{X_2})-R' \right|^+   \right]  \\
 &\dotleq \sum_{V\in\scV_n(\calY_2;Q_{UX_2})}\exp(-n D(V\|W_{Y_2|UX_2}|Q_{UX_2}) ) \exp\left[ -n  \left|   I( Q_{U|X_2},V|Q_{X_2})-R' \right|^+   \right] \\
 &\doteq\exp\left[-n \min_{V\in\scV_n(\calY_2;Q_{UX_2})}  \left( D(V\|W_{Y_2|UX_2}|Q_{UX_2}) + \left|   I( Q_{U|X_2},V|Q_{X_2})-R' \right|^+   \right)\right]
\end{align}
By appealing to continuity of exponents~\cite[Lem.~10.5]{Csi97} (minimum over conditional types is arbitrarily close to the minimum over conditional distributions for $n$ large enough), we obtain the exponent $F(R')$ in \eqref{eqn:F}.
 \end{proof}
\section{Compress-Forward (CF)}\label{sec:cf}

In this section, we state and prove an achievable error
exponent for CF. CF is   more complicated than PDF because the
relay does vector quantization on   the channel outputs $Y_2^n$ and forwards the description to
the destination. This quantized version of the channel output is denoted by $\hatY_2^n\in\hcalY_2^n$ and the error here is analyzed using covering  techniques Marton introduced for deriving the error exponent for rate-distortion~\cite{Marton74} and reviewed in Section~\ref{sec:review}.   Subsequently, the receiver decodes both the quantization  index  (a bin index associated with the quantized signal $\hatY_2^n$) and the message index.  This combination of covering and packing leads to a more involved analysis of the  error exponent that needs to leverage on ideas in Kelly-Wagner~\cite{Kel12} where the error exponent for Wyner-Ziv coding~\cite{wynerziv} was derived. It also leverages on a recently-developed proof  technique by Scarlett-Guill\'en i F\`abregas~\cite{Sca12} known as the {\em one-at-a-time union bound} given in \eqref{eqn:one_time}. This is particularly useful to analyze  the  error when two indices (of correlated signals) are to be  simultaneously decoded given a channel output.   At a   high level,  we operate on a conditional type-by-conditional type basis for the covering step at the relay. We also use an $\alpha$-decoding rule~\cite{CK81} for decoding the messages and the bin indices at the receiver.

This section is structured as follows: In Section~\ref{sec:cf_defs}, we provide basic definitions of the quantities that are used to state and prove the main theorem. The main theorem  is stated in Section~\ref{sec:cf_thm}. A detailed set of remarks to help in understanding the quantities involved in the main theorem is provided in Section~\ref{sec:cf_rmk}. Finally, the proof   is provided in Section~\ref{sec:cf_pf}. The notation for the codewords follows that in El Gamal and Kim~\cite[Thm.~16.4]{elgamal}.
\subsection{Basic Definitions} \label{sec:cf_defs}
Before we are able to state our result as succinctly as possible, we find it convenient to first define several quantities upfront. For CF, the following  types and conditional types will be kept fixed and hence can be optimized over eventually: input distributions $Q_{X_1} \in \scP_n(\calX_1)$, $Q_{X_2} \in \scP_n(\calX_2)$ and test channel $Q_{\hatY_2|Y_2 X_2} \in \scV_n(\hcalY_2; Q_{Y_2|X_2} Q_{X_2})$ for some (adversarial) channel realization $Q_{Y_2|X_2} \in \scV_n(\calY_2; Q_{X_2})$ to be specified later.

\subsubsection{Auxiliary Channels}
Let the  auxiliary channel  $W_{Q_{X_1}} :\calX_2\to\calY_2$ be defined as 
\begin{equation}
W_{Q_{X_1}} (y_2, y_3|x_2 ) := \sum_{x_1} W(y_2, y_3|x_1, x_2)Q_{X_1}(x_1) . \label{eqn:W1}
\end{equation} 
This is simply the  original relay channel averaged over $Q_{X_1}$. With a slight abuse of notation, we denote its marginals using the same notation, i.e., 
\begin{align}
W_{Q_{X_1}} ( y_3|x_2 ) &:=\sum_{y_2} W_{Q_{X_1}} (y_2, y_3|x_2 ) \label{eqn:WQx1_3} , \\
W_{Q_{X_1}} ( y_2|x_2 ) &:=\sum_{y_3} W_{Q_{X_1}} (y_2, y_3|x_2 )  \label{eqn:WQx1_2}.
\end{align}
Define another auxiliary channel  $W_{Q_{Y_2|X_2},Q_{\hatY_2|Y_2  X_2}}:\calX_1\times\calX_2\to\hcalY_2\times\calY_3$ as 
\begin{equation}
 W_{Q_{Y_2|X_2},Q_{\hatY_2|Y_2  X_2}}(\haty_2 , y_3| x_1, x_2) := \sum_{y_2} W(  y_3|x_1, x_2, y_2) Q_{\hatY_2|Y_2  X_2}(\haty_2|y_2, x_2)Q_{Y_2|X_2}(y_2|x_2)\label{eqn:W2}   
\end{equation}
where $W(  y_3|x_1, x_2, y_2)=W(y_2,  y_3|x_1, x_2)/ \sum_{y_3} W(y_2,  y_3|x_1, x_2)$ is the $\calY_3$-conditional distribution of the DM-RC. 
Note that $ W_{Q_{Y_2|X_2},Q_{\hatY_2|Y_2  X_2}}$ is simply the original relay channel averaged over both channel realization $Q_{Y_2|X_2}$ and test channel $Q_{\hatY_2|Y_2  X_2}$. Hence, if the realized conditional type of the relay input is $Q_{Y_2|X_2}$ and we fixed the test channel to be $Q_{\hatY_2|Y_2  X_2}$ (to be chosen dependent on $Q_{Y_2|X_2}$), then  we show that, effectively, the channel from $\calX_1\times\calX_2$ to $\hcalY_2\times\calY_3$ behaves as $W_{Q_{Y_2|X_2},Q_{\hatY_2|Y_2  X_2}}$. We make this precise in the proofs. See the steps leading to~\eqref{eqn:Pr_type_class}.

\subsubsection{Other Channels and Distributions}
For any two channels $Q_{Y_2|X_2},\tilQ_{Y_2|X_2}:\calX_2\to\calY_2$, define  two $\hcalY_2$-modified channels as follows:
\begin{align}
Q_{\hatY_2|X_2}(\haty_2|x_2) := \sum_{y_2} Q_{\hatY_2|Y_2  X_2}(\haty_2|y_2, x_2)Q_{Y_2|X_2}(y_2|x_2), \label{eqn:y2hatgivenx2} \\
\tilQ_{\hatY_2|X_2}(\haty_2|x_2) := \sum_{y_2} Q_{\hatY_2|Y_2  X_2}(\haty_2|y_2, x_2)\tilQ_{Y_2|X_2}(y_2|x_2).\label{eqn:tily2hatgivenx2}  
\end{align}
Implicit in these definitions are  $Q_{\hatY_2|Y_2  X_2}$, $Q_{Y_2| X_2}$ and $\tilQ_{Y_2| X_2}$ but these dependencies are suppressed for the sake of brevity. For any $V:\calX_1\times\calX_2\times\hcalY_2\to \calY_3$, let  the induced conditional distributions $ V _{Q_{X_1}}:\calX_2\times\hcalY_2\to\calY_3$ and $Q_{\hatY_2|X_2}  \times   V    :\calX_1\times\calX_2\to\hcalY_2\times\calY_3 $ be defined as 
\begin{align}
 V _{Q_{X_1}}(y_3|x_2, \haty_2)&:=\sum_{x_1}  V (y_3|x_1, x_2, \haty_2) Q_{X_1}(x_1) \label{eqn:VQX1} , \\*
( \tilQ_{\hatY_2|X_2} \times V  )    ( \haty_2 , y_3|x_1, x_2 )  &:=  V (y_3|x_1, x_2, \haty_2)\tilQ_{\hatY_2|X_2}(\haty_2|x_2). \label{eqn:QtimesV}
\end{align}

\subsubsection{Sets of Distributions and $\alpha$-Decoder}
Define the set of joint types $P_{X_1 X_2 \hatY_2  Y_3}$ with   marginals consistent with $Q_{X_1}, Q_{X_2}$ and $Q_{\hatY_2|X_2}$ as 
\begin{equation}
\scP_n(Q_{X_1}, Q_{X_2}, Q_{\hatY_2|X_2}):=\{P_{X_1 X_2 \hatY_2 Y_2}\!\in\!\scP_n(\calX_1\times\calX_2\times\hcalY_2\times\calY_3) \!: \!(P_{X_1}, P_{X_2},P_{\hatY_2|X_2 }) \! = \! ( Q_{X_1},  Q_{X_2},   Q_{\hatY_2|X_2 } ) \}. \label{eqn:setP}
\end{equation}
We will use the notation $\scP(Q_{X_1}, Q_{X_2}, Q_{\hatY_2|X_2})$ (without subscript $n$) to mean the same set as in \eqref{eqn:setP} without the restriction to types but all distributions in  $\scP(\calX_1\times\calX_2\times\hcalY_2\times\calY_3)$ satisfying the constraints in \eqref{eqn:setP}.
For any  four sequences $(x_1^n,x_2^n, \haty_2^n, y_3^n)$,   define the function $\alpha$ as 
\begin{equation}
\alpha(x_1^n, \haty_2^n, y_3^n|x_2^n)=\alpha(P,V) := D(V \| W_{Q_{Y_2|X_2},Q_{\hatY_2|Y_2  X_2}}|P) +  H(V|P) ,  \label{eqn:defq}
\end{equation}
where $P$ is the joint type of $(x_1^n, x_2^n,\haty_2^n)$, $V : \calX_1\times\calX_2\times\hcalY_2\to\calY_3$ is the conditional type of $ y_3^n$ given $(x_1^n, x_2^n,\haty_2^n)$, and $W_{Q_{Y_2|X_2},Q_{\hatY_2|Y_2  X_2}}$ is the channel defined in \eqref{eqn:W2}. Roughly speaking, to decode the bin index and message, we will maximize $\alpha$ over bin indices, messages and conditional types $Q_{Y_2|X_2}$.  This is exactly ML decoding~\cite{CK81} as discussed  in Section~\ref{sec:review}. Define the set of conditional types
\begin{align}
&\scK_n(  Q_{Y_2|X_2}, Q_{\hatY_2|Y_2  X_2} ) :=  \{V\in \scV_n(\calY_3; Q_{X_1} Q_{X_2} Q_{\hatY_2|X_2}) : \nn\\
&\qquad\qquad\qquad\alpha(Q_{X_1} Q_{X_2} Q_{\hatY_2|X_2} ,V )\ge\alpha(Q_{X_1} Q_{X_2} Q_{\hatY_2|X_2} , W_{Q_{Y_2|X_2},Q_{\hatY_2|Y_2  X_2}} ) \}. \label{eqn:setK}
\end{align}
Note that  $Q_{\hatY_2|X_2}$ is given in~\eqref{eqn:y2hatgivenx2} and is induced by the two arguments of $\scK_n$. Intuitively, the conditional types contained in $\scK_n(  Q_{Y_2|X_2}, Q_{\hatY_2|Y_2  X_2} )$ are those corresponding to sequences $y_3^n$ that lead to an error as  the likelihood computed with respect to $V$ is larger than (or  equal to) that for the true averaged channel $W_{Q_{Y_2|X_2},Q_{\hatY_2|Y_2  X_2}}$. The marginal types $Q_{X_1}, Q_{X_2}$ are fixed in the notation $\scK_n$ but we omit them for brevity.  We will use the notation $\scK(  Q_{Y_2|X_2}, Q_{\hatY_2|Y_2  X_2} )$ (without subscript $n$) to mean the same set as in \eqref{eqn:setK} without the restriction to  conditional types but all conditional distributions from $\calX_1\times\calX_2\times\hcalY_2$ to $\calY_3$ satisfying the constraints in \eqref{eqn:setK} so $\scK=\mathrm{cl}(\lim_{n\to\infty}\scK_n)$ where $\mathrm{cl}(\fndot)$ is the closure operation.

\subsection{Error Exponent for Compress-Forward}\label{sec:cf_thm}

\begin{theorem}[Error Exponent for Compress-Forward] \label{thm:cf}
Fix $b\in\bbN$ and ``Wyner-Ziv rate'' $R_2\ge 0$, distributions $Q_{X_1}\in\scP(\calX_1)$ and $Q_{X_2} \in \scP(\calX_2)$ and auxiliary alphabet $\hcalY_2$. The following is a lower bound on the reliability function:
\begin{align}
E(\Reff) \ge E_{\mathrm{cf}}^{(b)} (\Reff)  :=\frac{1}{b}\min\{ G_1(R, R_2), G_2(R, R_2) \}
\end{align}
where the constituent exponents are defined as 
\begin{align}
G_1(R, R_2) := \min_{V :\calX_2\to\calY_3} D(V \|  W_{Q_{X_1}} | Q_{X_2}) + \left|I(Q_{X_2}, V)-R_2 \right|^+ \label{eqn:G1cf}
\end{align}
where $W_{Q_{X_1}}$ is defined in \eqref{eqn:WQx1_3} and 
\begin{align}
G_2(R, R_2) := \min_{Q_{Y_2|X_2} : \calX_2\to\calY_2} D(Q_{Y_2|X_2} \|W_{Q_{X_1}}|Q_{X_2})+ \max_{Q_{\hatY_2|Y_2  X_2} : \calY_2\times\calX_2\to\hcalY_2} \, J(R, R_2, Q_{\hatY_2|Y_2  X_2}, Q_{Y_2|X_2}). \label{eqn:G2cf}
\end{align}
The quantity $J (R, R_2, Q_{\hatY_2|Y_2  X_2}, Q_{Y_2|X_2})$ that constitutes $G_2(R, R_2)$ is defined as 
\begin{align}
 & J (R, R_2, Q_{\hatY_2|Y_2  X_2}, Q_{Y_2|X_2})\nn\\
& := \min_{P_{X_1X_2\hatY_2 Y_3} \in \scP (Q_{X_1}, Q_{X_2}, Q_{\hatY_2|X_2}) } \Bigg\{D( P_{ \hatY_2 Y_3|X_1 X_2 } \, \| \,   W_{Q_{Y_2|X_2},Q_{\hatY_2|Y_2  X_2}} \, | \, Q_{X_1}Q_{X_2 } )  \nn\\
&   \qquad + \min_{   \tilQ_{Y_2|X_2} :\calX_2\to\calY_2}  \min_{    V \in \scK(  Q_{Y_2|X_2}, Q_{\hatY_2|Y_2  X_2} )  }\left[ \,\,\min_{l=1,2}\,\,  \psi_l( V ,\tilQ_{Y_2|X_2}, R, R_2, P_{X_1X_2\hatY_2 Y_3}) \right] \Bigg\}\label{eqn:Jcf}
\end{align} 
where $ W_{Q_{Y_2|X_2},Q_{\hatY_2|Y_2  X_2}},Q_{\hatY_2|X_2}$, and $\tilQ_{\hatY_2|X_2}$ are defined in~\eqref{eqn:W2}, \eqref{eqn:y2hatgivenx2} and~\eqref{eqn:tily2hatgivenx2}  respectively and the functions $\psi_l, l=1,2$ are defined as 
\begin{align}
\psi_1 ( V ,\tilQ_{Y_2|X_2}, R, R_2, P_{X_1X_2\hatY_2 Y_3}) := |I(Q_{X_1}, \tilQ_{\hatY_2|X_2}\times V|Q_{X_2})-R|^+ \label{eqn:psi1}
\end{align}
 and
\begin{align}  \label{eqn:theta_def}  
 \psi_2 ( V ,\tilQ_{Y_2|X_2},  R, R_2, P_{X_1X_2\hatY_2 Y_3}  ) := \left\{ \begin{array}{l}
 | \,  I(\tilQ_{\hatY_2|X_2},  V _{Q_{X_1}}  |Q_{X_2})    +  | I(Q_{X_1},  \tilQ_{\hatY_2|X_2}\times   V  |Q_{X_2}  )-R|^+     \\
\quad   + \, R_2- I (\tilQ_{Y_2|X_2},  Q_{\hatY_2|Y_2  X_2}|Q_{X_2})  \,   |^+ \\*
 \qquad\qquad\qquad\mbox{ \em if } \,\,\,  R_2\le I (\tilQ_{Y_2|X_2},  Q_{\hatY_2|Y_2  X_2}|Q_{X_2})  \\\\
    I(\tilQ_{\hatY_2|X_2}, V _{Q_{X_1}} |Q_{X_2})  +\,  |  I(Q_{X_1}, \tilQ_{\hatY_2|X_2}\times   V    |Q_{X_2})-R|^+        \\*
  \qquad\qquad\qquad\mbox{ \em  if } \,\,\,  R_2> I (\tilQ_{Y_2|X_2},  Q_{\hatY_2|Y_2  X_2}|Q_{X_2}) 
\end{array}\right. 
\end{align}
where $V_{Q_{X_1}}$ and $\tilQ_{\hatY_2|X_2}\times   V$ are defined in \eqref{eqn:VQX1} and \eqref{eqn:QtimesV} respectively.
\end{theorem}
Note that \eqref{eqn:theta_def} can be written more succinctly as 
\begin{align}
& \psi_2 ( V ,\tilQ_{Y_2|X_2},  R, R_2, P_{X_1X_2\hatY_2 Y_3}  ) \nn\\
&\quad  :=  \left| \,   | I(Q_{X_1},  \tilQ_{\hatY_2|X_2}\times   V   )-R|^+ +  I(\tilQ_{\hatY_2|X_2},  V _{Q_{X_1}}  |Q_{X_2}) - |  I (\tilQ_{Y_2|X_2},  Q_{\hatY_2|Y_2  X_2}|Q_{X_2})-R_2  \,   |^+ \, \right|^+. \label{eqn:succinct}
\end{align}

\subsection{Remarks on the Error Exponent for  Compress-Forward}\label{sec:cf_rmk}
In this Section, we dissect the  main features of the CF error exponent presented in Theorem~\ref{thm:cf}. To help the reader follow the proof, the point at which the various terms in the CF exponent are derived are summarized in Table~\ref{tab:cf}.

\begin{table}
\begin{center}
{ \renewcommand{\arraystretch}{1.6}
    \begin{tabular}{| c ||c| c |}
    \hline
    Expression  & Definition in & Derivation  in steps leading to \\    \hline
   $G_1(R, R_2)$  & \eqref{eqn:G1cf} &  \eqref{eqn:first_packing} \\     \hline
   $D(Q_{Y_2|X_2} \|W_{Q_{X_1}}|Q_{X_2})$ & \eqref{eqn:G2cf}  &  \eqref{eqn:prob_Y2} \\     \hline
   $D( P_{ \hatY_2 Y_3|X_1 X_2 } \, \| \,   W_{Q_{Y_2|X_2},Q_{\hatY_2|Y_2  X_2}} \, | \, Q_{X_1}Q_{X_2 } )$ & \eqref{eqn:Jcf} &   \eqref{eqn:Pr_type_class}  \\     \hline 
   $\psi_1 ( V ,\tilQ_{Y_2|X_2}, R, R_2, P_{X_1X_2\hatY_2 Y_3})$ & \eqref{eqn:psi1} &  \eqref{eqn:only_message}\\     \hline
First case in  $\psi_2 ( V ,\tilQ_{Y_2|X_2},  R, R_2, P_{X_1X_2\hatY_2 Y_3}  )$   & \eqref{eqn:theta_def}   & \eqref{eqn:gn2}   \\     \hline
Second case in $\psi_2 ( V ,\tilQ_{Y_2|X_2},  R, R_2, P_{X_1X_2\hatY_2 Y_3}  )$   & \eqref{eqn:theta_def}   & \eqref{eqn:gn1} and \eqref{eqn:gn12}  \\    \hline
    \end{tabular} }
\end{center}
\caption{Terms comprising $E_{\mathrm{cf}}^{(b)} (\Reff)$ and points of derivation in the proof in Section~\ref{sec:cf_pf}.}
\label{tab:cf}
\end{table}

\begin{enumerate}
\item We are free to choose the  independent input distributions $Q_{X_1}$ and $Q_{X_2}$, though these will be $n$-types for finite $n \in\bbN$.  We also have the freedom to choose any ``Wyner-Ziv rate'' $R_2\ge 0$. Thus, we can optimize over $Q_{X_1},Q_{X_2}$ and $R_2$. The $X_1$- and $X_2$-codewords are uniformly distributed in $\calT_{Q_{X_1}}$ and $\calT_{Q_{X_2}}$ respectively. 

\item As is well known  in CF~\cite{CEG}, the relay transmits a description $\haty_2^n(j)$ of its received sequence $y_2^n(j)$ (conditioned on $x_2^n(j)$  which is known to both relay and decoder) via a covering step. This explains the final mutual information term in~\eqref{eqn:theta_def}  which can be written as the rate loss $I(Y_2;\hatY_2|X_2)$, where   $(X_2, Y_2, \hatY_2)$ is distributed as $Q_{X_2}\times  \tilQ_{Y_2|X_2} \times  Q_{\hatY_2|Y_2  X_2}$.   Since covering results in   super-exponential decay in the error probability, this does not affect the overall exponent  since the {\em smallest} one dominates. See the steps leading to~\eqref{eqn:doubly} in the proof. 

\item  The exponent $G_1(R, R_2)$  in \eqref{eqn:G1cf} is analogous to $G(R')$  in \eqref{eqn:G1}. This represents the error exponent in the estimation of $X_2$'s index given $Y_3^n$ using MMI decoding.  However, in the CF proof, we do not use the packing lemma~\cite[Lem.~10.1]{Csi97}. Rather we construct a random code and show that on expectation (over the random code), the error probability decays exponentially fast with   exponent   $G_1(R, R_2)$.  This is the same as in Remark~\ref{item:random} in Section~\ref{sec:rmk_pdf} on the discussion of the PDF exponent.  In fact, other authors such as Kelly-Wagner~\cite{Kel12} and Moulin-Wang~\cite{Mou07} also derive their error exponents in this way.

\item  In the  exponent $G_2(R, R_2)$  in   \eqref{eqn:G2cf}, $Q_{Y_2|X_2}$ is the realization of the conditional type  of the received signal at the relay  $y_2^n(j)$ given $x_2^n(j)$. The divergence term $D(Q_{Y_2|X_2} \|W_{Q_{X_1}}|Q_{X_2})$ represents the deviation from the true channel behavior $W_{Q_{X_1}}$ similarly to the interpretation of the random coding error exponent for point-to-point channel coding in \eqref{eqn:rcee}. We can optimize for the  conditional distribution  (test channel) $Q_{\hatY_2|Y_2  X_2}$ compatible with $Q_{Y_2|X_2}Q_{X_2}$. This explains the outer minimization over $Q_{Y_2|X_2}$  and inner maximization over $Q_{\hatY_2|Y_2  X_2}$ in~\eqref{eqn:G2cf}. This is  a game-theoretic-type result along the same lines as in~\cite{Kel12, Mou07, Dar11}.

\item The first part of $J$ given by $\psi_1$ in \eqref{eqn:psi1}   represents the incorrect decoding of the index of $X_1^n$ (message $M_j$)  as well as the conditional type $Q_{Y_2|X_2}$  given that the bin index of  the description $\hatY_2^n$ is decoded correctly.   The second part of   $J$ given by $\psi_2$ in \eqref{eqn:theta_def} represents the incorrect decoding  the bin index of  $\hatY_2^n$,  the index of $X_1^n$ (message $M_j$) as well as the conditional type $Q_{Y_2|X_2}$. We see the different sources of ``errors'' in~\eqref{eqn:Jcf}: There is  a  minimization over the different types of channel behavior  represented by $P_{X_1X_2\hatY_2 Y_3}$  and also a minimization over estimated conditional types $\tilQ_{Y_2|X_2}$. Subsequently, the error involved in $\alpha$-decoding of the message and the bin index of the description sequence is represented by the minimization over $V \in  \scK(  Q_{Y_2|X_2}, Q_{\hatY_2|Y_2  X_2} )$.

 \item We see that the freedom of choice of the ``Wyner-Ziv rate'' $R_2\ge 0$ allows us to operate in one of two distinct regimes.  This can be seen from the two different cases in~\eqref{eqn:theta_def}.  The number of description codewords in $\hcalY_2^n$ is designed to be $\exp(nI (\tilQ_{Y_2|X_2},  Q_{\hatY_2|Y_2  X_2}|Q_{X_2}) ) = \exp(n I(Y_2;\hatY_2|X_2))$ to first order in the exponent, where the choice of  $Q_{\hatY_2|Y_2  X_2}$ depends on the realized conditional type $Q_{Y_2|X_2}$. The number of Wyner-Ziv bins is $\exp(nR_2)$. When $R_2\le   I (\tilQ_{Y_2|X_2},  Q_{\hatY_2|Y_2  X_2}|Q_{X_2})$,  we do additional  Wyner-Ziv binning as there are more description sequences than bins. If $R_2$ is larger than $I (\tilQ_{Y_2|X_2},  Q_{\hatY_2|Y_2  X_2}|Q_{X_2})$, no additional   binning is required. The {\em excess Wyner-Ziv   rate} is thus
 \begin{equation}
 \Delta R_2   := I (\tilQ_{Y_2|X_2},  Q_{\hatY_2|Y_2  X_2}|Q_{X_2})-R_2  .\label{eqn:excess_wz}
 \end{equation}
 This explains the presence of this term in \eqref{eqn:theta_def}   or equivalently, \eqref{eqn:succinct}.

\item \label{item:multi} For the analysis of  the error in decoding    the bin and message indices, if we simply apply  the packing lemmas in~\cite{Csi97, Csis00, Kor80b, CKM, Har08, CK81}, this would result in a suboptimal rate vis-\`a-vis CF.  This is because   the conditional correlation of $X_1$ and $\hatY_2$ given $X_2$ would not be  taken into account.    Thus, we need to analyze this error exponent more carefully  using the one-at-a-time union bound~\cite{Sca12} used originally for the multiple-access channel under mismatched decoding and stated in \eqref{eqn:one_time}. Note that the sum of the first two mutual informations  (ignoring the $|\fndot|^+$)  in~\eqref{eqn:theta_def}, which represents the bound on the sum of the message rate and the description sequence rate (cf.~\cite[Eq.~(16.11)]{elgamal}), can be written as
\begin{align}
& I(\hatY_2;Y_3|X_2)+ I( X_1;\hatY_2,Y_3|X_2)   \label{eqn:entropy0}\\
  &\quad = H(X_1|X_2) +  H(\hatY_2|X_2) + H(Y_3|X_2)      -  H(X_1 , \hatY_2,  Y_3|X_2) , \label{eqn:entropy} 
\end{align}
where $(X_1,X_2, \hatY_2,Y_3)\sim Q_{X_1}\cp  Q_{X_2}\cp  \tilQ_{\hatY_2|X_2} \cp V$. The entropies in~\eqref{eqn:entropy} demonstrate  the  symmetry between $X_1$ and $\hatY_2$ when they are decoded {\em jointly} at the receiver $Y_3$. The expressions in \eqref{eqn:entropy0}--\eqref{eqn:entropy} are in fact the {\em conditional mutual  information among three variables} $I(X_1;\hatY_2; Y_3|X_2)$ defined, for example,  in Liu and Hughes~\cite[Sec.~III.B]{Liu96}. This quantity is also called the {\em multi-information} by Studen\'{y} and Vejnarov\'{a}~\cite{studeny}.  
 
 In addition, the proof shows that by modifying the  order of applying union bounds in \eqref{eqn:one_time}, we can  get another achievable   exponent. Indeed, $\psi_2$ in~\eqref{eqn:succinct} can be strengthened to be the maximum of the expression on its right-hand-side and 
\begin{align}
 \psi_2' ( V ,\tilQ_{Y_2|X_2},  R, R_2, P_{X_1X_2\hatY_2 Y_3}  )  :=  \left|   I(Q_{X_1},  \tilQ_{\hatY_2|X_2}\times   V   )-R +  \left| I(\tilQ_{\hatY_2|X_2},  V _{Q_{X_1}}  |Q_{X_2})     - \Delta R_2  \, \right|^+\, \right|^+ ,\label{eqn:succinct2}
\end{align}
where $\Delta R_2 $ is defined in \eqref{eqn:excess_wz}.
 
 \item \label{item:cf_pos} From the exponents in Theorem~\ref{thm:cf}, it is clear upon eliminating $R_2$ (if $R_2$ is chosen small enough so that Wyner-Ziv binning is necessary) that we recover the CF lower bound in \eqref{eqn:cf}.  Indeed,  if  $\psi_1$ is active in the minimization in \eqref{eqn:Jcf}, the first term in \eqref{eqn:cf} is positive if and only if the error exponent $G_2$   is positive  for some choice of  distributions $Q_{X_1}$,  $Q_{X_2}$ and   $Q_{\hatY_2|Y_2  X_2}$.  Also, if $\psi_2$ is active  in the minimization in \eqref{eqn:Jcf}  and  $R_2$ is chosen sufficiently small  (so that the first clause of \eqref{eqn:theta_def} is active), $G_2$  is positive  if 
\begin{align}
R &< R_2 + I(X_1;\hatY_2 Y_3|X_2) + I(\hatY_2; Y_3|X_2) - I(\hatY_2; Y_2|X_2) \label{eqn:cf_just} \\
&< I(X_2;Y_3) + I(X_1;\hatY_2 Y_3|X_2) + I(\hatY_2; Y_3|X_2) - I(\hatY_2; Y_2|X_2 )  \label{eqn:useR2}\\
&= I(X_1X_2;Y_3) + I(X_1;\hatY_2 |X_2 Y_3) + I(\hatY_2; Y_3|X_2) - I(\hatY_2; Y_2|X_2 )  - I(X_1 Y_3; \hatY_2 | X_2Y_2) \label{eqn:cf2}\\
&= I(X_1X_2;Y_3) + I(X_1 Y_3;\hatY_2 |X_2)  - I(X_1 Y_2 Y_3; \hatY_2| X_2)  \label{eqn:cf4}\\
&= I(X_1X_2;Y_3) - I(Y_2; \hatY_2 | X_1 X_2 Y_3) .  \label{eqn:cf3}
\end{align}
for some  $Q_{X_1}$,  $Q_{X_2}$ and   $Q_{\hatY_2|Y_2  X_2}$. In~\eqref{eqn:useR2}, we used the fact that $G_1$ in~\eqref{eqn:G1cf} is positive if and only if  $R_2<I(X_2;Y_3)$ and in \eqref{eqn:cf2} we also used the chain rule for mutual information (twice) and the Markov chain $\hatY_2-(X_2, Y_2)-(X_1, Y_3)$ \cite[pp.\ 402]{elgamal} so the final mutual information term $I(X_1 Y_3; \hatY_2 | X_2, Y_2)=0$.  Equations~\eqref{eqn:cf4} and \eqref{eqn:cf3} follow by repeated applications of the chain rule for mutual information. Equation~\eqref{eqn:cf3} matches the second term in   \eqref{eqn:cf}.  
 \item Lastly, the evaluation of the CF exponent appears to be extremely difficult because of (i) non-convexity of the optimization problem and (ii)   multiple nested optimizations. It is, however, not apparent how to simplify the CF exponent (to the Gallager form, for example) to make it amendable to evaluation  for a given DM-RC~$W$. 
 \end{enumerate}
\subsection{Proof of Theorem~\ref{thm:cf}}\label{sec:cf_pf}
 \begin{proof}
{\bf Random Codebook Generation:} We again use block-Markov coding~\cite{CEG}. Fix types $Q_{X_1} \in\scP_n(\calX_1)$ and $Q_{X_2}\in\scP_n(\calX_2)$ as well as rates $R,R_2\ge 0$. For each $j\in [b]$, generate a random codebook in the following manner. Randomly and independently generate $\exp(nR)$ codewords $x_1^n(m_j)\sim \Unif[\calT_{Q_{X_1}}]$, where $\Unif[\calA]$ is the uniform distribution over the  finite set $\calA$.   Randomly and independently generate $\exp(n R_2 )$ codewords $x_2^n(l_{j-1})\sim \Unif[\calT_{Q_{X_2}}]$.  Now for every  $Q_{Y_2|X_2}\in\scV_n(\calY_2; Q_{X_2})$ fix a  different test channel $Q_{\hatY_2|Y_2  X_2} (Q_{Y_2|X_2})\in \scV_n(\hcalY_2;Q_{Y_2|X_2} Q_{X_2})$. For  every $Q_{Y_2|X_2} \in  \scV_n(\calY_2;  Q_{X_2})$ and every $x_2^n(l_{j-1})$ construct a   conditional type-dependent codebook $\calB( Q_{Y_2|X_2}, l_{j-1}) \subset\hcalY_2^n$   of integer size   $|  \calB( Q_{Y_2|X_2}, l_{j-1}) |$ whose rate, which we call the {\em inflated rate}, satisfies
\begin{align}
\tilR_2(Q_{Y_2|X_2})& :=\frac{1}{n} \log |  \calB( Q_{Y_2|X_2}, l_{j-1}) |   = I( Q_{Y_2|X_2},Q_{\hatY_2|Y_2  X_2} (Q_{Y_2|X_2})|Q_{X_2}) + \nu_n  ,  \label{eqn:choiceB}
\end{align}
where $\nu_n\in \Theta(\frac{\log n}{n})$ and more precisely,
\begin{equation}
 \frac {(|\calX_2|  |\calY_2|  |\hcalY_2| +2)\log (n+1)}{n}\le \nu_n  \le \frac {(|\calX_2|  |\calY_2|  |\hcalY_2| +3)\log (n+1)}{n}. \label{eqn:nu}
\end{equation}
Each sequence    in $\calB( Q_{Y_2|X_2}, l_{j-1})$ is indexed as $\haty_2^n(k_j|l_{j-1})$ and is drawn independently according to  the uniform distribution $\Unif[ \calT_{Q_{\hatY_2|X_2}  }(x_2^n(l_{j-1}) ) ]$ where $Q_{\hatY_2|X_2} $ is the marginal induced by $Q_{Y_2|X_2}$ and $Q_{\hatY_2|Y_2  X_2} (Q_{Y_2|X_2})$.  See the definition in \eqref{eqn:y2hatgivenx2}. Depending on the choice of $R_2$, do one of the following:
\begin{itemize}
\item  If $R_2 \le \tilR_2(Q_{Y_2|X_2})$,   partition  the  conditional type-dependent codebook $\calB( Q_{Y_2|X_2}, l_{j-1})$ into $\exp(nR_2)$ equal-sized bins  $\calB_{l_j}( Q_{Y_2|X_2}, l_{j-1})$, $l_j\in  [\exp(n  R_2 )]$. 
\item If $R_2 > \tilR_2(Q_{Y_2|X_2})$,  assign each element of $\calB( Q_{Y_2|X_2}, l_{j-1})$ a unique index in $[\exp(n  R_2 )]$. 
\end{itemize}

{\bf Transmitter Encoding:}  The encoder transmits $x_1^n(m_j)$ at block $ j\in[b]$. 

{\bf Relay Encoding:}  At the end of block $j\in [b]$, the relay encoder has $x_2^n (l_{j-1})$ (by convention $l_0 :=1$) and its input sequence $y_2^n(j)$. It computes the conditional type $Q_{Y_2|X_2}\in\scV_n(\calY_2; Q_{X_2})$.  Then it searches in $\calB( Q_{Y_2|X_2}, l_{j-1})$ for a description sequence
\begin{equation}
\haty_2^n(\hatk_j|l_{j-1}) \in\calT_{ Q_{\hatY_2|Y_2  X_2} (Q_{Y_2|X_2} )} (y_2^n(j), x_2^n (l_{j-1}) ). \label{eqn:covering_step}
\end{equation}
If more than one such sequence exists, choose one uniformly at random   in $\calB( Q_{Y_2|X_2}, l_{j-1})$  from those satisfying~\eqref{eqn:covering_step}. If none exists, choose   uniformly at random from $\calB( Q_{Y_2|X_2}, l_{j-1})$. Identify the bin index $\hatl_j$ of $\haty_2^n(\hatk_j|l_{j-1})$ and send $x_2^n(\hatl_{j})$ in block $j+1$. 

{\bf Decoding:}
At the end of block $j+1$,  the receiver has  the channel output $y_3^n(j+1)$. It does MMI decoding~\cite{Goppa} by finding $\hatl_j$ satisfying 
\begin{equation}
\hatl_j :=\argmax_{l_j\in [\exp(nR_2)]}\,\,  \hatI( x_2^n (l_j)\wedge y_3^n(j+1) ). \label{eqn:ljhat}
\end{equation}
Having identified $\hatl_{j-1},\hatl_{j }$ from~\eqref{eqn:ljhat}, find message $\hatm_j$,  index $\hatk_j$ and conditional type $\hatQ_{Y_2|X_2}^{(j)} \in\scV_n(\calY_2 ; Q_{X_2})$ satisfying
\begin{align} 
(\hatm_j, \hatk_j , \hatQ_{Y_2|X_2}^{(j)}) = \argmax_{( m_j, k_j, Q_{Y_2|X_2}): \haty_2^n(k_j|\hatl_{j-1}) \in \calB_{\hatl_j}( Q_{Y_2|X_2} ,  \hatl_{j-1} )} 
\alpha  \left( x_1^n(m_j) , \haty_2^n(k_j|\hatl_{j-1}), y_3^n(j ) \,\Big|\, x_2^n(\hatl_{j-1})  \right)  , \label{eqn:mjkj} 
\end{align} 
where the function $\alpha$ was defined in \eqref{eqn:defq}. This  is an $\alpha$-decoder~\cite{CK81} which finds  the $(\hatm_j, \hatk_j, \hatQ_{Y_2|X_2}^{(j)} )$  maximizing $\alpha$  subject to  $\haty_2^n(k_j|\hatl_{j-1}) \in \calB_{\hatl_j}( \hatQ_{Y_2|X_2}^{(j)} , \hatl_{j-1} )$, where $\hatl_{j-1},\hatl_{j}$ were found in~\eqref{eqn:ljhat}.  
We decode $Q_{Y_2|X_2}$ so as to know which bin $\hat{y}_2^n(k_j | \hat{l}_{j-1})$ lies in. Since there are only polynomially many conditional types $Q_{Y_2|X_2}$ this does not degrade our CF error exponent. The decoding of the conditional type is inspired partly by Moulin-Wang's derivation of the error exponent for Gel'fand-Pinsker coding~\cite[pp.~1338]{Mou07}.
Declare that $\hatm_j$ was sent.

Let us pause to understand why we used two different rules  (MMI and   ML-decoding) in \eqref{eqn:ljhat} and \eqref{eqn:mjkj}. In the former, we are simply decoding a single index $l_j$ given $y_3^n(j+1)$ hence MMI suffices. In the latter, we are decoding two indices $m_j$ and $k_j$  and as mentioned in Section~\ref{sec:main_contr}, we need to take into account the conditional correlation between $X_1$ and $\hatY_2$. If we had simply used an MMI decoder, it appears that  we would have a strictly smaller quantity $I(\hatY_2;Y_3|X_2) + I(X_1; Y_3|X_2\hatY_2) = I(X_1 \hatY_2; Y_3  |X_2)$  in the analogue of \eqref{eqn:entropy0}--\eqref{eqn:entropy}, which represents the upper bound on the sum of $R$ and the excess Wyner-Ziv  rate $\Delta R_2$ defined in \eqref{eqn:excess_wz}.  This   would not recover the CF lower bound in the steps from~\eqref{eqn:cf_just}--\eqref{eqn:cf3}. Hence, we used the ML-decoder in \eqref{eqn:mjkj}.

{\bf Analysis of Error Probability:}  We now analyze the error probability. Assume as usual that $M_j=1$ for all $j\in [b-1]$ and let $L_{j-1}, L_j$ and $K_j$ be indices chosen by the relay in block $j$.  First, note that as in~\eqref{eqn:union_bd},
\begin{equation}
\bbP(\hatM\ne M)\le  (b-1)  \left( \epsilon_{\rmR} + 2 \epsilon_{\rmD,1} +   \epsilon_{\rmD,2}   \right) ,\label{eqn:ub_cf}
\end{equation}
where $\epsilon_{\rmR}$ is the error event that there is no description sequence  $\haty_2^n(\hatk_j|l_{j-1})$ in the   bin $\calB( Q_{Y_2|X_2}, l_{j-1})$ that satisfies~\eqref{eqn:covering_step} (covering error), 
\begin{equation}
\epsilon_{\rmD,1}:=\bbP( \hatL_{j}\ne L_j)
\end{equation}
is the error probability in decoding the wrong $l_j$ bin index,  and
\begin{equation}
 \epsilon_{\rmD,2}:=\bbP(  \hatM_j  \ne  1 |L_j , L_{j-1} \mbox{ decoded correctly})
\end{equation} 
 is the error probability in decoding   the message    incorrectly.  See the proof of compress-forward in~\cite[Thm.~16.4]{elgamal} for details of the calculation in~\eqref{eqn:ub_cf}. Again, since $b$ is a constant, it does not affect the exponential dependence  on the error probability in \eqref{eqn:ub_cf}. We bound each error probability separately.  Note that the error probability is an average over the random codebook generation so, by the usual random coding argument, as long as this average is small, there must exist  at least one code with such a small error probability.


{\bf Covering Error $\epsilon_\rmR$:} For $\epsilon_{\rmR}$, we follow the proof idea in \cite[Lem.\ 2]{Kel12}. In the following, we let the conditional type of $y_2^n$ given $x_2^n$ be $Q_{Y_2|X_2}$. Consider the conditional covering error conditioned on $X_2^n=x_2^n$ and $Y_2^n=y_2^n$, namely  
\begin{align}
\epsilon_{\rmR}(x_2^n,y_2^n)  := \bbP(\calF  |Y_2^n=y_2^n , X_2^n=x_2^n ), \label{eqn:cover_err}
\end{align}
where $\calF$ is the event that {\em every} sequence $\haty_2^n(k_j|l_{j-1})\in \calB(Q_{Y_2|X_2},l_{j-1})$ does {\em not} satisfy \eqref{eqn:covering_step}. Let $\exp_{\rme}(t) :=\rme^t$. ($\rme$ is the base of the natural logarithm.)  Now we use the mutual independence of the codewords in $\calB(Q_{Y_2|X_2}, l_{j-1} )$ and basic properties of types (Lemmas~\ref{lem:types} and~\ref{lem:joint_typ}) to assert that
\begin{align}
\epsilon_{\rmR}(x_2^n,y_2^n)   
 & =\prod_{k_j  } \bbP \left(\hatY_2^n(k_j|l_{j-1}) \notin \calT_{ Q_{\hatY_2|Y_2  X_2} (Q_{Y_2|X_2} )} (  y_2^n, x_2^n  ) \right) \label{eqn:prod} \\
& =  \left[ 1- \bbP \left(\hatY_2^n(1|l_{j-1}) \in \calT_{ Q_{\hatY_2|Y_2  X_2} (Q_{Y_2|X_2} )} (  y_2^n, x_2^n  )  \right)\right]^{| \calB(Q_{Y_2|X_2}, l_{j-1} )|}\\
&\le  \left[ 1- (n+1)^{-|\calX_2||\calY_2| |\hcalY_2| }  \exp  (-n   I (Q_{Y_2|X_2},  Q_{\hatY_2|Y_2  X_2}(Q_{Y_2|X_2})|Q_{X_2}   )   \right]^{| \calB(Q_{Y_2|X_2}, l_{j-1} )|}\\
 &\le \exp_{\rme}\left[   - (n+1)^{-|\calX_2||\calY_2| |\hcalY_2| } \exp(-n [ I (Q_{Y_2|X_2},  Q_{\hatY_2|Y_2  X_2}(Q_{Y_2|X_2})|Q_{X_2} )   - \tilR_2(Q_{Y_2|X_2}  )] )       \right] \label{eqn:exp_ineq} \\
&\le\exp_{\rme}\left[-(n+1)^2 \right],\label{eqn:doubly}
\end{align}
where the product in~\eqref{eqn:prod} extends over all indices $k_j$ for which $\haty_2^n(k_j|l_{j-1}) \in \calB(Q_{Y_2|X_2},l_{j-1})$  for some fixed realization of $l_{j-1}$ which we can condition on. Inequality  \eqref{eqn:exp_ineq} follows from the inequality $(1-x)^k\le\exp_{\rme}(-kx)$, \eqref{eqn:doubly} follows from the choice of  $\tilR_2(Q_{Y_2|X_2} )$ and $\nu_n$ in~\eqref{eqn:choiceB} and~\eqref{eqn:nu} respectively. This derivation is   similar to the  type covering lemma for source coding with a fidelity criterion (rate-distortion) with excess distortion probability  by Marton~\cite{Marton74}. Now, let $\calE_{\rmC}$ be the set of all pairs $(x_2^n,y_2^n)$ that lead to a covering error according to the decoding rule \eqref{eqn:covering_step}. We follow the argument proposed by Kelly-Wagner~\cite[pp.~5100]{Kel12} to assert that
\begin{align}
\epsilon_{\rmR} & = \sum_{(x_2^n, y_2^n) \in \calE_{\rmC}} \bbP( X_2^n = x_2^n , Y_2^n = y_2^n)  \label{eqn:emR} \\
& = \sum_{(x_2^n, y_2^n) \in \calE_{\rmC}} \bbP( X_2^n = x_2^n , Y_2^n = y_2^n,\calF) \\
& \le \sum_{(x_2^n, y_2^n) \in \calE_{\rmC}} \bbP(\calF| X_2^n = x_2^n , Y_2^n = y_2^n) \\
& = \sum_{(x_2^n, y_2^n) \in \calE_{\rmC}}\epsilon_{\rmR}(x_2^n,y_2^n)   \label{eqn:use_F} \\
& \le \sum_{(x_2^n, y_2^n) \in \calE_{\rmC}} \exp_{\rme}\left[-(n+1)^2 \right]\doteq 0 . \label{eqn:emR_end} 
\end{align}
The punchline is that    $\epsilon_{\rmR}$ decays super-exponentially (i.e., $\epsilon_\rmR \doteq 0$ or the exponent is infinity) and thus it does not affect the  overall error exponent since the {\em smallest} one dominates. 

{\bf  First Packing Error $\epsilon_{\rmD,1}$:} We assume $L_j=1$ here. The calculation here is very similar to that in Section~\ref{prf:pdf}  but we provide the details for completeness. We evaluate  $\epsilon_{\rmD,1}$ by partitioning the sample space into subsets where $X_2^n(1)$ takes on various values $x_2^n \in \calT_{Q_{X_2}}$. Thus, we have
\begin{align}
\epsilon_{\rmD,1} &\le\bbP\left( \exists\, \till_j \ne 1: \hatI ( X_2^n(\till_j) \wedge Y_3^n(j+1)) \ge \hatI ( X_2^n(1) \wedge Y_3^n(j+1)) \right) \\
&=\sum_{x_2^n \in \calT_{Q_{X_2}}}\frac{1}{ |  \calT_{Q_{X_2}}|}\beta_n(x_2^n) 
\end{align}
where 
\begin{equation}
\beta_n(x_2^n) :=\bbP\left( \exists\, \till_j \ne 1: \hatI ( X_2^n(\till_j) \wedge Y_3^n(j+1)) \ge \hatI ( X_2^n(1) \wedge Y_3^n(j+1)) \,\big|\, X_2^n(1)=x_2^n\right) .
\end{equation}
It can easily be seen that $\beta_n(x_2^n)$ is independent of $x_2^n \in \calT_{Q_{X_2}}$ so we abbreviate $\beta_n(x_2^n)$ as $\beta_n$. Because $X_1^n(1)$ is generated uniformly at random from  $\calT_{Q_{X_1}}$, we have 
\begin{align}
\beta_n = \sum_{x_1^n \in \calT_{Q_{X_1}}}\frac{1}{ |  \calT_{Q_{X_1}}|} \sum_{y_3^n } W^n(y_3^n|x_1^n, x_2^n) \mu_n(y_3^n)
\end{align}
where 
\begin{equation}
\mu_n (y_3^n):= \bbP\left( \exists\, \till_j \ne 1: \hatI ( X_2^n(\till_j) \wedge Y_3^n(j+1)) \ge \hatI ( X_2^n(1) \wedge Y_3^n(j+1)) \,\big|\, Y_3^n(j+1)=y_3^n,X_2^n(1)=x_2^n\right) . \label{eqn:def_mu} 
\end{equation}
Note that the event before the conditioning in $\mu_n(y_3^n)$ does not depend on the event $\{X_1^n(1)=x_1^n\}$ so we drop the dependence on $x_1^n$  from the notation  $\mu_n(y_3^n)$. Also the only source of randomness in the probability in \eqref{eqn:def_mu} is $X_2^n(\till_j), \till_j\ne 1$ which is independent of $X_1^n(1)$.  We continue to bound $\beta_n$ as follows:
\begin{align}
\beta_n& \le  \sum_{x_1^n \in \calT_{Q_{X_1}}}(n+1)^{|\calX_1|} \exp(-nH(Q_{X_1}))\sum_{y_3^n } W^n(y_3^n|x_1^n, x_2^n) \mu_n(y_3^n) \label{eqn:lower_bd_tc}\\
&=\sum_{x_1^n \in \calT_{Q_{X_1}}}(n+1)^{|\calX_1|}Q_{X_1}^n(x_1^n)\sum_{y_3^n } W^n(y_3^n|x_1^n, x_2^n) \mu_n(y_3^n) \label{eqn:lower_bd_tc1}\\
& \le (n+1)^{|\calX_1|} \sum_{y_3^n } \mu_n(y_3^n)\sum_{x_1^n }Q_{X_1}^n(x_1^n) W^n(y_3^n|x_1^n, x_2^n)  \\
& = (n+1)^{|\calX_1|}   \sum_{y_3^n } W_{Q_{X_1}}^n(y_3^n|  x_2^n) \mu_n(y_3^n) \label{eqn:use_def_WQX1} .
\end{align}
where \eqref{eqn:lower_bd_tc} follows from the lower bound on the size of a type class (Lemma~\ref{lem:types}), \eqref{eqn:lower_bd_tc1} follows from the fact that the $Q_{X_1}^n$-probability of a sequence $x_1^n$ of type $Q_{X_1}$ is exactly $ \exp(-nH(Q_{X_1}))$  and \eqref{eqn:use_def_WQX1} is an application of the definition of $W_{Q_{X_1}}$ in \eqref{eqn:WQx1_3}.   It remains to bound  $\mu_n (y_3^n)$ in \eqref{eqn:def_mu}. We do so by first applying the union bound
\begin{align}
\mu_n (y_3^n)\le\min\left\{ 1, \exp(nR_2)\tau_n(y_3^n) \right\} \label{eqn:bound_mu} ,
\end{align}
where 
\begin{equation}
\tau_n(y_3^n):=\bbP\left( \hatI ( X_2^n(2) \wedge Y_3^n(j+1)) \ge \hatI ( X_2^n(1) \wedge Y_3^n(j+1)) \,\big|\, Y_3^n(j+1)=y_3^n,X_2^n(1)=x_2^n\right).
\end{equation}
We now use notation $\tilV:\calY_3\to\calX_2$ to denote a {\em reverse channel}. Also let $P_{y_3^n}$ be the type of $y_3^n$. Let $\scR(Q_{X_2})$ be the class of reverse channels satisfying $\sum_{y_3}\tilV(x_2|y_3)P_{y_3^n}(y_3)=Q_{X_2}(x_2)$. Then, we have 
\begin{equation}
\tau_n(y_3^n)= \sum_{\substack{ \tilV \in\scV_n(\calX_2;P_{y_3^n})\cap\scR(Q_{X_2}) : \\ \hatI(x_2^n\wedge y_3^n)\le I(P_{y_3^n},\tilV)}}
\bbP\left(X_2^n \in  \calT_{\tilV}(y_3^n) \right), \label{eqn:tau_n}
\end{equation}
where $X_2^n$ is uniformly distributed over the type class $\calT_{Q_{X_2}}$. From Lemma~\ref{lem:joint_typ2} (with the identifications $\calX_1\leftarrow \emptyset$, $V\leftarrow Q_{X_2}$,  $V'\leftarrow P_{y_3^n}$,  $W\leftarrow\tilV$), we have that for every $\tilV\in\scR(Q_{X_2})$,
\begin{equation}
\bbP\left(X_2^n \in  \calT_{\tilV}(y_3^n) \right)\dotleq\exp(-n I(P_{y_3^n},\tilV)).
\end{equation}
Hence using the clause in~\eqref{eqn:tau_n} yields
\begin{align}
\tau_n(y_3^n) 
&\dotleq   \exp(-n\hatI(x_2^n\wedge y_3^n)).
\end{align}
 Substituting this into the the bound for $ \mu_n (y_3^n)$ in \eqref{eqn:bound_mu} yields
\begin{equation}
\mu_n (y_3^n)\dotleq\exp\left[-n | \hatI(x_2^n\wedge y_3^n)-R_2|^+\right].
\end{equation}
Plugging this back into the bound for $\beta_n$ in  \eqref{eqn:use_def_WQX1} yields
\begin{align}
\beta_n & \dotleq \sum_{y_3^n } W_{Q_{X_1}}^n(y_3^n|  x_2^n)\exp\left[-n | \hatI(x_2^n\wedge y_3^n)-R_2|^+\right]\\
& =\sum_{V\in\scV_n(\calY_3;Q_{X_2}) } W_{Q_{X_1}}^n(\calT_{V}(x_2^n)|  x_2^n)\exp\left[-n | I(Q_{X_2},V)-R_2|^+\right]\\
&\le\sum_{V\in\scV_n(\calY_3;Q_{X_2}) } \exp \left[-n \left(  D(V\|W_{Q_{X_1}}|Q_{X_2})+ | I(Q_{X_2},V)-R_2|^+   \right)\right]. \label{eqn:first_packing}
\end{align}
This gives the exponent $G_1(R,R_2)$ in \eqref{eqn:G1cf} upon minimizing over all  $V\in\scV_n(\calY_3;Q_{X_2})$.

{\bf  Second Packing Error $\epsilon_{\rmD,2}$:} We evaluate $\epsilon_{\rmD,2}$ by  partitioning the sample space into subsets where the conditional type of relay input $y_2^n$ given relay   output $x_2^n$ is $Q_{Y_2|X_2} \in\scV_n(\calY_2; Q_{X_2})$. That is, 
\begin{equation}
\epsilon_{\rmD,2} = \sum_{Q_{Y_2|X_2}\in\scV_n(\calY_2 ; Q_{X_2})} \bbP(Y_2^n\in\calT_{Q_{Y_2|X_2} }(X_2^n)) \varphi_n( Q_{Y_2|X_2})  \label{eqn:ed3_ub} 
\end{equation}
where  $\varphi_n( Q_{Y_2|X_2})$  is defined as 
\begin{equation}
\varphi_n( Q_{Y_2|X_2}) :=\bbP\left( \hatM_j\ne 1  \,\Big|\, L_j,L_{j-1}   \mbox{ decoded correctly} , Y_2^n\in\calT_{Q_{Y_2|X_2} }(X_2^n) \right) . \label{eqn:defF}
\end{equation}
We bound the probability in \eqref{eqn:ed3_ub} and $\varphi_n( Q_{Y_2|X_2})$ in the following. Then we optimize over all conditional types  $Q_{Y_2|X_2}\in\scV_n(\calY_2 ; Q_{X_2})$. This corresponds to the minimization in \eqref{eqn:G2cf}.

The probability in \eqref{eqn:ed3_ub} can be first bounded using the same steps  in~\eqref{eqn:lower_bd_tc} to~\eqref{eqn:use_def_WQX1} as
\begin{align}
\bbP   (Y_2^n  \in\calT_{Q_{Y_2|X_2} }(X_2^n)) 
&\le (n+1)^{|\calX_1|}\sum_{x_2^n\in\calT_{Q_{X_2}}}\frac{1}{|\calT_{Q_{X_2}}|}\sum_{y_2^n \in \calT_{Q_{Y_2|X_2}}(x_2^n)} W_{Q_{X_1}}^n(y_2^n|x_2^n) \label{eqn:prob_type1}
\end{align}
Now by using Lemma~\ref{lem:types},
\begin{align}
\bbP   (Y_2^n  \in\calT_{Q_{Y_2|X_2} }(X_2^n)) 
&\dotleq \exp \left[ -n D( Q_{Y_2|X_2} \| W_{Q_{X_1}} | Q_{X_2}) \right]\label{eqn:prob_Y2}.
\end{align}
This gives the first part of the exponent $G_2(R, R_2)$ in  \eqref{eqn:G2cf}.


Recall the notations $\scP_n(Q_{X_1}, Q_{X_2}, Q_{\hatY_2|X_2})$ and  $\scK_n(  Q_{Y_2|X_2}, Q_{\hatY_2|Y_2  X_2} )$ in \eqref{eqn:setP} and \eqref{eqn:setK} respectively. These sets will be used in the subsequent calculation.  Implicit in the calculations below is the fact that $Y_2^n \in \calT_{Q_{Y_2|X_2}}(X_2^n)$ and also for the fixed $Q_{Y_2|X_2}$ we have the fixed test channel $Q_{\hatY_2|Y_2  X_2}(Q_{Y_2|X_2})$ which we will denote more succinctly as $Q_{\hatY_2|Y_2  X_2}$. See~\eqref{eqn:G2cf} and the codebook generation. In the following steps, we simply use the notation $\calT_{P_{X_1 X_2 \hatY_2 Y_3}}$ as an abbreviation for the event that the random quadruple of sequences $(X_1^n(1), X_2^n(L_{j-1}), \hatY_2^n(K_j|L_{j-1}), Y_3^n(j) )$ belongs to the type class $\calT_{P_{X_1 X_2 \hatY_2 Y_3}}$. Following the one-at-a-time union bound strategy in~\cite{Sca12} (see \eqref{eqn:one_time}), we now bound $\varphi_n( Q_{Y_2|X_2})$ in~\eqref{eqn:defF} by conditioning on various joint types $P_{X_1 X_2 \hatY_2 Y_3} \in \scP_n(Q_{X_1}, Q_{X_2}, Q_{\hatY_2|X_2})$
\begin{align}
\varphi_n( Q_{Y_2|X_2})\le \sum_{P_{X_1 X_2 \hatY_2 Y_3} \in \scP_n(Q_{X_1}, Q_{X_2}, Q_{\hatY_2|X_2})} \bbP \left( \calT_{P_{X_1 X_2 \hatY_2 Y_3}}  \right) \bbP\left( \bigcup_{ V  \in \scK_n(  Q_{Y_2|X_2}, Q_{\hatY_2|Y_2  X_2} ) }\calE_{ V } \,\bigg|\,  \calT_{P_{X_1 X_2 \hatY_2 Y_3}}\right) \label{eqn:split_varphi} ,
\end{align}
where $ Q_{\hatY_2|X_2}$ is specified in \eqref{eqn:y2hatgivenx2} based on  $Q_{\hatY_2|Y_2  X_2}$  and $Q_{Y_2|X_2}$   and   the event $\calE_{ V }$ is defined as 
\begin{equation}
\calE_{ V } := \bigcup_{\tilQ_{Y_2|X_2}  \in  \scV_n(\calY_2; Q_{X_2}  )  } \calE_{ V }   (\tilQ_{Y_2|X_2})\label{eqn:Ev} , \end{equation}
with the constituent events defined as 
\begin{align}
 \calE_{ V }   (\tilQ_{Y_2|X_2}) :=\bigcup_{\tilm_j \in [\exp(nR)]\setminus\{1\}}\bigcup_{\tilk_j \in \calB_{ \hatL_j } ( \tilQ_{Y_2 | X_2 }, \hatL_{j-1}) } \calE_{ V }(\tilQ_{Y_2|X_2} ,\tilm_j,\tilk_j)  \label{eqn:EvQ} ,
 \end{align}
 and
 \begin{align}
\calE_{ V }(\tilQ_{Y_2|X_2},\tilm_j,\tilk_j)  :=\left\{   \left(X_1^n(\tilm_j) , X_2^n(L_{j-1}), \hatY_2^n(\tilk_j|L_{j-1}), Y_3^n(j  ) \right) \in \calT_{Q_{X_1}   Q_{X_2} \tilQ_{\hatY_2|X_2}   V } \right\}\label{eqn:EvQmk}  . 
\end{align}
Recall the definition of $\tilQ_{\hatY_2|X_2} $  in \eqref{eqn:tily2hatgivenx2}. This is a function of the decoded conditional type $\tilQ_{Y_2|X_2} $. Note that $\tilQ_{Y_2|X_2}$ indexes a decoded conditional type (of which there are only polynomially many), $\tilm_j$ indexes an incorrect decoded message and $\tilk_j$ indexes a  correctly ($\tilk_j=K_j$) or incorrectly decoded  bin index ($\tilk_j\ne K_j$). The union over $\tilk_j$ extends over the entire bin  $\calB_{ \hatL_j } ( \tilQ_{Y_2 | X_2 }, \hatL_{j-1})$ and not only over incorrect bin indices. This is because an error is declared only if $\tilm_j\ne 1$. Essentially, in the crucial step in \eqref{eqn:split_varphi}, we have conditioned on the channel behavior  (from $\calX_1\times\calX_2$ to $\hcalY_2\times\calY_3$) and identified the set of conditional types (indexed by $V$) that leads to an error based on the $\alpha$-decoding step in~\eqref{eqn:mjkj}.

Now we will bound the constituent elements in \eqref{eqn:split_varphi}.  Recall that   $X_1^n$ is drawn uniformly at random from $\calT_{Q_{X_1}}$, $X_2^n$ is drawn uniformly at random from $\calT_{Q_{X_2}}$  and  $\hatY_2^n$ is drawn uniformly at random from $\calT_{Q_{\hatY_2|Y_2  X_2}}(y_2^n, x_2^n)$  where $y_2^n\in\calT_{Q_{Y_2|X_2}}(x_2^n)$ and $Q_{Y_2|X_2}$ is the conditional type fixed in \eqref{eqn:ed3_ub}.  Note that if we are given that $Y_2^n\in\calT_{Q_{Y_2|X_2}}(x_2^n)$, it must be  uniformly distributed in $\calT_{Q_{Y_2|X_2}}(x_2^n)$ given $X_2^n = x_2^n$.  Finally,  $Y_3^n$ is drawn from  the relay channel $W^n(\fndot|y_{2}^n,x_{1}^n, x_{2}^n)$. Using these facts, we can establish that the  first probability in  \eqref{eqn:split_varphi} can be expressed as
\begin{equation}  
\bbP \left( \calT_{P_{X_1 X_2 \hatY_2 Y_3}}  \right) = \sum_{x_1^n \in \calT_{Q_{X_1}}}  \sum_{x_2^n \in \calT_{Q_{X_2}}} \frac{ \vartheta(x_1^n, x_2^n)}{| \calT_{Q_{X_1}}| | \calT_{Q_{X_2}}|}, \label{eqn:prob_TP} 
\end{equation}
where 
\begin{equation}
\vartheta(x_1^n, x_2^n) := \sum_{y_2^n \in \calT_{Q_{Y_2|X_2}}(x_2^n)}\frac{1}{| \calT_{Q_{Y_2|X_2}}(x_2^n)|} \sum_{ \substack{ (\haty_2^n, y_3^n) \in \calT_{P_{\hatY_2 Y_3|X_1 X_2} } (x_1^n, x_2^n)\\  \haty_2^n \in \calT_{Q_{\hatY_2|Y_2X_2}}(y_2^n,x_2^n)}} \frac{W^n(y_3^n|y_2^n, x_1^n, x_2^n)}{ | \calT_{Q_{\hatY_2|Y_2X_2}}(y_2^n,x_2^n) |} . \label{eqn:vartheta_def}
\end{equation}
Notice that the term $| \calT_{Q_{Y_2|X_2}}(x_2^n)|^{-1} \bone\{y_2^n \in   \calT_{Q_{Y_2|X_2}}(x_2^n) \}$ in~\eqref{eqn:vartheta_def} indicates that $Y_2^n$ is  uniformly distributed in  $ \calT_{Q_{Y_2|X_2}}(x_2^n)$ given that $X_2^n=x_2^n$.
We now bound $\vartheta(x_1^n, x_2^n)$ by the same logic as the steps from~\eqref{eqn:lower_bd_tc} to~\eqref{eqn:use_def_WQX1} (relating size of shells to probabilities of sequences). More precisely, 
\begin{align}
\vartheta(x_1^n, x_2^n) &\le (n+1)^{ |\calX_2| |\calY_2| ( 1+ |\hcalY_2|)} \sum_{y_2^n \in \calT_{Q_{Y_2|X_2}}(x_2^n)}Q_{Y_2|X_2}^n(y_2^n|x_2^n)  \,  \times \nn\\
 &\qquad\times \sum_{ \substack{ (\haty_2^n, y_3^n) \in \calT_{P_{\hatY_2 Y_3|X_1 X_2} } (x_1^n, x_2^n)\\  \haty_2^n \in \calT_{Q_{\hatY_2|Y_2X_2}}(y_2^n,x_2^n)}} Q_{\hatY_2|Y_2X_2}^n(\haty_2^n|y_2^n, x_2^n)W^n(y_3^n|y_2^n, x_1^n, x_2^n) \\
  &\dotleq    \sum_{   (\haty_2^n, y_3^n) \in \calT_{P_{\hatY_2 Y_3|X_1 X_2} } (x_1^n, x_2^n)}  \sum_{y_2^n} Q_{Y_2|X_2}^n(y_2^n|x_2^n)   Q_{\hatY_2|Y_2X_2}^n(\haty_2^n|y_2^n, x_2^n)W^n(y_3^n|y_2^n, x_1^n, x_2^n)  \label{eqn:use_def_WQ1}\\
   &=    \sum_{ (\haty_2^n, y_3^n) \in \calT_{P_{\hatY_2 Y_3|X_1 X_2} } (x_1^n, x_2^n) }  W_{Q_{Y_2|X_2},Q_{\hatY_2|Y_2  X_2}}^n(\haty_2^n , y_3^n| x_1^n, x_2^n)  \label{eqn:use_def_WQ}\\
 &\doteq \exp\left[-n D( P_{\hatY_2   Y_3|X_1 X_2} \, \| \,  W_{Q_{Y_2|X_2},Q_{\hatY_2|Y_2  X_2}}  \, | \, Q_{X_1}Q_{X_2}  ) \right] \label{eqn:bound_h}
\end{align}
where \eqref{eqn:use_def_WQ1} follows by dropping the constraints  $y_2^n \in \calT_{Q_{Y_2|X_2}}(x_2^n)$ and $\haty_2^n \in \calT_{Q_{\hatY_2|Y_2X_2}}(y_2^n,x_2^n)$ and reorganizing the sums,  \eqref{eqn:use_def_WQ} follows from the definition of  $ W_{Q_{Y_2|X_2},Q_{\hatY_2|Y_2  X_2}}$ in \eqref{eqn:W2} and~\eqref{eqn:bound_h} follows by Lemma~\ref{lem:types}. Substituting \eqref{eqn:bound_h} into  \eqref{eqn:prob_TP} yields the exponential bound
\begin{equation}
\bbP \left( \calT_{P_{X_1 X_2 \hatY_2 Y_3}}  \right) \dotleq\exp\left[ -n D( P_{\hatY_2   Y_3|X_1 X_2} \, \| \,  W_{Q_{Y_2|X_2},Q_{\hatY_2|Y_2  X_2}}  \, | \, Q_{X_1}Q_{X_2}  ) \right] \label{eqn:Pr_type_class} .
\end{equation}
This gives the first part of the expression $J(R, R_, Q_{\hatY_2 | Y_2 X_2} , Q_{Y_2|X_2})$ in \eqref{eqn:Jcf}.

Hence, all that remains is to bound the second probability (of the union) in \eqref{eqn:split_varphi}. 
We first deal with the case where the decoded bin index $\tilk_j$ is correct, i.e., equal to $K_j$.  In this case,   the $X_1^n$ codeword  is conditionally independent of the outputs $(\hat{Y}_2^n , Y_3^n)$ given $X_2^n$. This is because the index of $X_1^n(\tilde{m}_j)$ is not equal to $1$ (i.e., $\tilde{m}_j\ne 1$) and the index of $\hat{Y}_2^n$ is decoded correctly. Thus we can view $(\hat{Y}_2^n , Y_3^n)$ as the outputs of a channel with input $X_1^n(1)$ and side-information available at the decoder $X_2^n$ \cite[Eq.~(7.2)]{elgamal}.  By the definition of $\calE_{ V }(\tilQ_{Y_2|X_2} ,\tilm_j,\tilk_j)$ in \eqref{eqn:EvQmk},
\begin{align}
\bbP\left(\calE_{ V }(\tilQ_{Y_2|X_2} ,\tilm_j,K_j)  \,\Big|\, \calT_{P_{X_1 X_2 \hatY_2 Y_3}}\right)=\bbP\left( (\bar{\haty}_2^n, \bary_3^n)\in\calT_{Q_{\hatY_2|X_2}\times V}(X_1^n(\tilm_j), X_2^n(L_{j-1})) \,\Big|\, \calT_{P_{X_1 X_2 \hatY_2 Y_3}} \right), \label{eqn:cond_inde}
\end{align}
where we have used the bar notation $(\bar{\haty}_2^n, \bary_3^n)$ to denote an arbitrary pair of sequences in the ``marginal shell'' induced by  $\barW(\haty_2,y_3|x_1,x_2):=\sum_{x_1}Q_{X_1}(x_1)  \tilQ_{\hatY_2|X_2}(\haty_2|x_2) V(y_3|x_1,x_2,\haty_2)$. We can condition on any realization of $X_2^n(L_{j-1}) =x^n_2\in \calT_{Q_{X_2}}$ here. Lemma~\ref{lem:joint_typ} (with identifications $P\leftarrow Q_{X_2}$, $V \leftarrow   Q_{X_1}$,   $V'\leftarrow \tilQ_{\hatY_2|X_2}\times V$, and $\barW\leftarrow W$) yields,
\begin{align}
\bbP\left(\calE_{ V }(\tilQ_{Y_2|X_2} ,\tilm_j,K_j)  \,\Big|\, \calT_{P_{X_1 X_2 \hatY_2 Y_3}}\right)\doteq\exp\left[ -n I(Q_{X_1}, \tilQ_{\hatY_2|X_2}\times V|Q_{X_2}) \right],  \label{eqn:use_joint_typ}
\end{align}
and so by applying the union bound (and using the fact that probability cannot exceed one),
\begin{equation}
\bbP\left(\bigcup_{\tilm_j \in [\exp(nR)]\setminus\{1\}}  \calE_{ V }(\tilQ_{Y_2|X_2} ,\tilm_j,K_j) \,\bigg|\,  \calT_{P_{X_1 X_2 \hatY_2 Y_3}} \right)\dotleq\exp\left[ -n \left|I(Q_{X_1}, \tilQ_{\hatY_2|X_2}\times V|Q_{X_2})-R   \right|^+\right]. \label{eqn:only_message}
\end{equation}
 This corresponds to the case involving $\psi_1$ in \eqref{eqn:psi1}. 

For the other case  (i.e., $\psi_2$ in \eqref{eqn:theta_def}) where both the message and bin index  are incorrect ($\tilm_j\ne 1$ and $\tilk_j\ne K_j$), slightly more intricate analysis is required.  For any conditional type $\tilQ_{Y_2|X_2}$,  define the {\em excess Wyner-Ziv rate given $\tilQ_{Y_2|X_2}$} as
\begin{equation}
\Delta R_2  (\tilQ_{Y_2|X_2}) := \tilR_2 (\tilQ_{Y_2|X_2})-R_2 \label{eqn:excess}
\end{equation}
where the inflated rate $\tilR_2 (\tilQ_{Y_2|X_2})$ is defined in~\eqref{eqn:choiceB}. This is exactly  \eqref{eqn:excess_wz} but we make the dependence on $\tilQ_{Y_2|X_2}$ explicit in \eqref{eqn:excess}. Assume for the moment that  $\Delta R_2  (\tilQ_{Y_2|X_2}) \ge 0$. Equivalently, this means that $R_2\le I (\tilQ_{Y_2|X_2},  Q_{\hatY_2|Y_2  X_2}|Q_{X_2})+\nu_n$, which is, up to the $\nu_n \in \Theta(\frac{\log n}{n})$ term, the first clause in~\eqref{eqn:theta_def}. Again, using bars to denote random variables generated uniformly from their respective marginal type classes and arbitrary sequences in their respective marginal type classes, define as in~\cite{Sca12}
\begin{align}
\xi_n ( V  , \tilQ_{Y_2|X_2})&:= \exp(n\Delta R_2  (\tilQ_{Y_2|X_2})) \cdot \bbP\left[ (\barx_2^n, \bar{\hatY}_2^n, y_3^n) \in\calT_{Q_{X_2} \tilQ_{\hatY_2|X_2}   V _{Q_{X_1}}  } \right]  \label{eqn:xin}\\*
\zeta_n ( V  , \tilQ_{Y_2|X_2}) &:= \exp(n R) \cdot\bbP\left[ (\barx_2^n, \bar{\haty}_2^n,\barX_1^n, \bary_3^n) \in \calT_{Q_{X_2} Q_{X_1} (\tilQ_{\hatY_2|X_2}   \times   V )}\right]\label{eqn:zetan}
\end{align}
where the conditional distributions $V_{Q_{X_1}}$ and  $\tilQ_{\hatY_2|X_2}   \times   V$ are defined in \eqref{eqn:VQX1} and \eqref{eqn:QtimesV} respectively. By applying the one-at-a-time union to the two unions in \eqref{eqn:EvQ} as was done in~\cite{Sca12} (see \eqref{eqn:one_time} in the Introduction), we obtain
\begin{align}
\bbP\left[ \calE_{ V }   (\tilQ_{Y_2|X_2})  \,\big|\,  \calT_{P_{X_1 X_2 \hatY_2 Y_3}}  \right] \le \gamma_n (V,\tilQ_{Y_2|X_2}) \label{eqn:perr_EV}
\end{align}  
where 
\begin{equation} 
\gamma_n (V,\tilQ_{Y_2|X_2}):=\min \left\{1, \xi_n ( V  , \tilQ_{Y_2|X_2})\cdot\min\{1,\zeta_n( V  , \tilQ_{Y_2|X_2})\} \right\} .\label{eqn:g_def}
\end{equation}
Now by using  the same reasoning that led to \eqref{eqn:only_message} (i.e., Lemma~\ref{lem:joint_typ}), we see that \eqref{eqn:xin} and \eqref{eqn:zetan} evaluate to
\begin{align}
\xi_n( V  , \tilQ_{Y_2|X_2}) & \doteq \exp\left[ -n \left( I(\tilQ_{\hatY_2|X_2 } ,   V _{Q_{X_1}} | Q_{X_2} ) -\Delta R_2  ( \tilQ_{Y_2|X_2}) \right) \right] \label{eqn:eval_xi} \\
\zeta_n ( V  , \tilQ_{Y_2|X_2})& \doteq \exp\left[-n \left( I(Q_{X_1},    \tilQ_{\hatY_2|X_2}\times   V    |Q_{X_2})   -R \right) \right] \label{eqn:eval_zeta}
\end{align}
Hence, $\gamma_n (V,\tilQ_{Y_2|X_2})$ in \eqref{eqn:g_def} has the following exponential behavior: 
\begin{align}
\gamma_n (V,\tilQ_{Y_2|X_2})  \doteq \exp\left[    -n   \left| I(\tilQ_{\hatY_2|X_2 } ,   V _{Q_{X_1}} | Q_{X_2} )  \!- \!\Delta R_2  ( \tilQ_{Y_2|X_2}) +  \left|  I(Q_{X_1},    \tilQ_{\hatY_2|X_2}\times   V    |Q_{X_2})   -R  \right|^+ \right|^+ \right]. \label{eqn:gn1}
\end{align}
Note that we can swap the order of the union bounds in the bounding of the probability in~\eqref{eqn:perr_EV}. As such, the probability in of the event $\calE_{ V }   (\tilQ_{Y_2|X_2})$ can also be upper bounded by 
\begin{equation}
\gamma_n' (V,\tilQ_{Y_2|X_2}):=\min \left\{1,\zeta_n( V  , \tilQ_{Y_2|X_2}) \cdot\min\{1,  \xi_n ( V  , \tilQ_{Y_2|X_2}) \} \right\} ,\label{eqn:g_def2}
\end{equation}
which, in view of \eqref{eqn:eval_xi} and~\eqref{eqn:eval_zeta}, has the exponential behavior
\begin{align}
\gamma_n' (V,\tilQ_{Y_2|X_2})  \doteq \exp\left[    -n   \left|   I(Q_{X_1},    \tilQ_{\hatY_2|X_2}\times   V    |Q_{X_2})   -R + \left| I(\tilQ_{\hatY_2|X_2 } ,   V _{Q_{X_1}} | Q_{X_2} )  \!- \!\Delta R_2  ( \tilQ_{Y_2|X_2}) \right|^+   \right|^+ \right]. \label{eqn:gn12}
\end{align}
Compare and contrast \eqref{eqn:gn12} to \eqref{eqn:gn1}.

Now consider $\Delta R_2  (\tilQ_{Y_2|X_2}) <0$.  Equivalently, this means that $R_2 >I (\tilQ_{Y_2|X_2},  Q_{\hatY_2|Y_2  X_2}|Q_{X_2})+\nu_n$, which is, up to the $\nu_n\in \Theta(\frac{\log n}{n})$ term in \eqref{eqn:nu}, the second clause in \eqref{eqn:theta_def}. In this case, we simply upper bound $\exp(n\Delta R_2  (\tilQ_{Y_2|X_2}))$ by unity and hence, $\xi_n ( V  , \tilQ_{Y_2|X_2})$ as 
\begin{equation}
\xi_n ( V  , \tilQ_{Y_2|X_2}) \dotleq \exp\left[ -n  I(\tilQ_{\hatY_2|X_2 } ,   V _{Q_{X_1}} | Q_{X_2} )  \right]
\end{equation}
and this yields
\begin{align}
\gamma_n (V,\tilQ_{Y_2|X_2}) \dotleq   \exp\left[   -n   \left( I(\tilQ_{\hatY_2|X_2 } ,   V _{Q_{X_1}} | Q_{X_2} )   +  \left|  I(Q_{X_1},    \tilQ_{\hatY_2|X_2}\times   V    |Q_{X_2})   -R  \right|^+  \right) \right]. \label{eqn:gn2}
\end{align}
Uniting the definition of $\tilR_2(Q_{Y_2|X_2})$ in~\eqref{eqn:choiceB}, the probabilities in~\eqref{eqn:prob_Y2} and~\eqref{eqn:Pr_type_class}, the case where only the message is incorrect in~\eqref{eqn:only_message}, the definition of the excess Wyner-Ziv  rate  $\Delta R_2  (\tilQ_{Y_2|X_2})$ in~\eqref{eqn:excess} and the case where both message and bin index are incorrect in~\eqref{eqn:gn1} and~\eqref{eqn:gn2} yields the exponent $G_2(R,R_2)$ in~\eqref{eqn:G2cf} as desired. 

Finally, we remark that the alternative exponent given in \eqref{eqn:succinct2} comes from using \eqref{eqn:gn12} instead of \eqref{eqn:gn1}.
\end{proof}

\section{An Upper Bound for the Reliability Function} \label{sec:sp}
\newcommand{\cmac}{\mathrm{C-MAC}}
\newcommand{\cbc}{\mathrm{C-BC}}
In this section, we state and prove an upper bound  on the reliability function per Definition~\ref{def:er}. This bound is inspired by  Haroutunian's exponent for  channels with  feedback~\cite{Har77}. Also see \cite[Ex.~10.36]{Csi97}. The upper bound on the reliability function is stated in Section~\ref{sec:sphere_pack}, discussions are provided in Section~\ref{sec:rem_sphere_pack} and the proof is detailed in Section~\ref{sec:prf_sphere_pack}.

\subsection{The Upper Bound on the Reliability Function}\label{sec:sphere_pack}

Before  we state the upper bound, define the function 
\begin{equation}
\rvC_{\mathrm{cs}}(V):=  \max_{P_{X_1X_2} \in\scP(\calX_1\times\calX_2)} \, \min\left\{I(P_{X_1X_2}, V_{Y_3|X_1X_2}) ,  I(P_{X_1|X_2}, V|P_{X_2}) \right\}, \label{eqn:cutset}
\end{equation}
where $V$ represents a  transition matrix from $\calX_1\times\calX_2$ to $\calY_2\times \calY_3$ and $V_{Y_3|X_1X_2}$ is its $\calY_3$-marginal. We recognize that \eqref{eqn:cutset} is the cutset upper bound on all achievable rates for the DM-RC $V$ (introduced in \eqref{eqn:cutset_intro}) but written in a different form in which the distributions are explicit. Note that the  subscript $\mathrm{cs}$ stands for {\em cutset}.  

\begin{theorem}[Upper Bound on the Reliability Function] \label{thm:upper}
We have the following upper bound on the reliability function:
\begin{align}
E(R)\le E_{\mathrm{cs}}(R) &:=\min_{ \substack{ V : \calX_1\times\calX_2\to\calY_2\times \calY_3 \\ \rvC_{\mathrm{cs}}(V)\le R}} \, \max_{P_{X_1X_2}\in\scP(\calX_1\times\calX_2)} \, D( V\| W| P_{X_1X_2} )  . \label{eqn:ub_rel}
\end{align}
\end{theorem}
The proof of this Theorem is provided in Section~\ref{sec:prf_sphere_pack}.
\subsection{Remarks on the Upper Bound on the Reliability Function}\label{sec:rem_sphere_pack}
\begin{enumerate}

\item  \label{item:pos_sp} Clearly, the upper bound $E_{\mathrm{cs}}(R)$ is positive if and only if $R< \rvC_{\mathrm{cs}}(W)$.    Furthermore, because $P_{X_1 X_2}\mapsto D(V\|W |P_{X_1 X_2})$ is linear,  the maximum is achieved at a particular symbol pair $(x_1, x_2)$, i.e., 
\begin{equation}
E_{\mathrm{cs}}(R)  :=\min_{ \substack{ V : \calX_1\times\calX_2\to\calY_2\times \calY_3 \\ \rvC_{\mathrm{cs}}(V)\le R}} \,  \max_{x_1, x_2 } \, D( V(\fndot, \fndot|x_1, x_2) \, \| \, W(\fndot, \fndot|x_1, x_2) ).  \label{eqn:ub_rel4}
\end{equation}
The computation of the $E_{\mathrm{cs}}(R)$ appears to be less challenging than CF but is still difficult because   $\rvC_{\mathrm{cs}}(V)$ is not convex in general and so $E_{\mathrm{cs}}(R)$ in \eqref{eqn:ub_rel} or \eqref{eqn:ub_rel4} are not convex optimization problems. Finding the joint distribution $P_{X_1 X_2}$ that achieves  $\rvC_{\mathrm{cs}}(V)$ for any $V$ is also not straightforward as one needs to develop an extension of the Blahut-Arimoto  algorithm~\cite[Ch.~8]{Csi97}. We do not explore this further   as developing efficient numerical algorithms is  not the focus of the current work.

\item \label{item:loose} We expect that, even though the cutoff rate (rate at which the exponent transitions from being positive to  being zero) is the cutset bound, the upper bound we derived in \eqref{eqn:ub_rel} is quite   loose relative to the achievability  bounds prescribed by Theorems~\ref{thm:pdf} and~\ref{thm:cf}. This is because  the  achievability theorems leverage on block-Markov coding and hence the achievable  exponents are attenuated by the number of blocks  $b$ causing significant delay when $b$ is large. (See Section~\ref{sec:num} for a numerical example.)  This  factor is not present in Theorem~\ref{thm:upper}. 


\item One method to  strengthen the exponent is to consider  a more restrictive class of codes,  namely codes with {\em finite memory}. For this class of codes, there exists some integer $l\ge 1$ (that does not depend on $n$) such that $g_i(y_2^{i-1}) = g_i (y_{2, i-l}^{i-1})$ for all $i\in [n]$. Under a similar  assumption for the discrete memoryless channel (DMC), Como and Nakibo\u{g}lu~\cite{Como} showed that the sphere-packing bound~\cite{Haroutunian68} is an upper bound for DMCs with feedback, thus improving on Haroutunian's original result~\cite{Har77}. In our setting, this would mean that 
\begin{equation}
\tilE_{\mathrm{cs}}(R) := \max_{P_{X_1X_2}\in\scP(\calX_1\times\calX_2)} \min_{ \substack{ V : \calX_1\times\calX_2\to\calY_2\times \calY_3  \\  \min\left\{I(P_{X_1X_2}, V_{Y_3|X_1X_2}) ,  I(P_{X_1|X_2}, V|P_{X_2}) \right\}\le R}} D(V\| W|P_{X_1X_2})
\end{equation}
is also an  upper bound on the reliability function. This bound, reminiscent of the sphere-packing exponent~\cite{Haroutunian68} is, in general, tighter (smaller) than \eqref{eqn:ub_rel} for general DM-RCs $W$.  We defer this  extension to future work.

\item \label{item:sphere} To prove an upper bound on the reliability function for channel coding problems, many authors first demonstrate  a strong converse. See  the proof      that the sphere-packing exponent is an upper bound for the reliability function of a DMC without feedback  in  Csisz\'ar-K\"orner~\cite[Thm.~10.3]{Csi97}; the proof of the sphere-packing exponent for asymmetric broadcast channels in~\cite[Thm.~1(b) or Eq.~(12)]{Kor80b}; and the proof of  the upper bound of the reliability function for Gel'fand-Pinsker coding~\cite[Thms.~2 and 3]{tyagi} for example.  Haroutunian's original proof of the former does not require the strong converse  though~\cite[Eq.~(26)]{Haroutunian68}. For relay channels and relay networks, the strong converse above the cutset bound  was recently  proved by Behboodi and Piantanida~\cite{Beh11, Beh12} using information spectrum methods~\cite{Han10} but  we do not need the strong converse for the proof of Theorem~\ref{thm:upper}.  Instead we leverage on a more straightforward change-of-measure technique by Palaiyanur~\cite{Palai11}.   
\end{enumerate}
\subsection{Proof of Theorem~\ref{thm:upper}}\label{sec:prf_sphere_pack}
\begin{proof}
Fix $\delta>0$  and let a given  DM-RC  $V: \calX_1\times\calX_2\to\calY_2\times \calY_3$  be such that $\rvC_{\mathrm{cs}}(V) \le R-\delta$.  Since the rate $R$ is larger than the cutset bound  $\rvC_{\mathrm{cs}}(V)$,   by the weak converse for DM-RCs~\cite[Thm.~16.1]{elgamal}, the average error probability assuming the DM-RC is   $V$ (defined in Definition~\ref{def:err_prb}) is bounded away from zero, i.e.,  
\begin{equation}
\rmP_{\rme}(V)\ge \eta , \label{eqn:Vge}
\end{equation}
for some $\eta>0$ that depends only on $R$.  Because   the signal $Y_2^{i-1}$  is provided to the relay encoder for each time $i \in [n]$ and we do not have a strong converse statement in~\eqref{eqn:Vge}, we cannot apply the change-of-measure technique in Csisz\'ar-K\"orner~\cite[Thm.~10.3]{Csi97}. We instead follow the proof strategy proposed  in Palaiyanur's thesis~\cite[Lem.~18]{Palai11} for channels with feedback. 
 
First, an additional bit of notation: For any message $m\in\calM$,   joint input type $P\in\scP_n(\calX_1 \times\calX_2)$, conditional type $U\in\scV_n(\calY_2\times\calY_3;P)$ and code $(f,g^n,\varphi)$, define the {\em relay shell} as follows  
\begin{equation}
\calA(m, P, U) := \left\{(y_2^n, y_3^n) : (f(m), g^n(y_2^n)) \in\calT_P ,  (y_2^n, y_3^n )\in\calT_U(f(m), g^n(y_2^n)) \right\}. \label{eqn:defB}
\end{equation}
Note that $(f(m), g^n(y_2^n) ) \in\calX_1^n\times \calX_2^n$ can be regarded as the channel input when the input to the  relay node is $y_2^n$. So $\calA(m,P,U)$ is the set of all $(y_2^n, y_3^n)$ that lie in the $U$-shell of the channel inputs $(x_1^n, x_2^n)$ which are of  joint type $P$ and these channel inputs result from sending message $m$. 

From the definition of $\rmP_{\rme}(W)$ in \eqref{eqn:def_err_prob}, we have
\begin{align}
\rmP_{\rme}(W) = \frac{1}{|\calM|}\sum_{m\in\calM} \sum_{ \substack{ P\in\scP_n(\calX_1\times\calX_2) \\  U \in\scV_n(\calY_2\times\calY_3;P)} } \sum_{\substack{ (y_2^n,y_3^n) \in\calA(m, P, U) \\ y_3^n \in\calD_m^c}}  \bbP_W((Y_2^n,Y_3^n) =(y_2^n,y_3^n)| M=m) ,
\end{align}
where  we partitioned $\calX_1^n\times\calX_2^n$ into sequences of the same type $P$ and we partitioned $\calY_2^n\times\calY_3^n$ into conditional types $U$ compatible with $P$.   Now, we change the measure in the inner probability to $V$ as follows:
\begin{align}
\rmP_{\rme}(W)&= \frac{1}{|\calM|}\sum_{m\in\calM} \sum_{ \substack{ P\in\scP_n(\calX_1\times\calX_2) \\  U \in\scV_n(\calY_2\times\calY_3;P)} } \sum_{\substack{ (y_2^n,y_3^n) \in\calA(m, P, U) \\ y_3^n \in\calD_m^c}} \bbP_V((Y_2^n,Y_3^n) =(y_2^n,y_3^n)| M=m) \nn\\
&\qquad\qquad\qquad \times \prod_{i=1}^n \frac{W(y_{2i}, y_{3i} | f_i(m), g_i(y_2^{i-1})) }{V(y_{2i}, y_{3i} | f_i(m), g_i(y_2^{i-1})) } \label{eqn:change_me}\\
&= \frac{1}{|\calM|}\sum_{m\in\calM} \sum_{ \substack{ P\in\scP_n(\calX_1\times\calX_2) \\  U \in\scV_n(\calY_2\times\calY_3;P)} } \sum_{\substack{ (y_2^n,y_3^n) \in\calA(m, P, U) \\ y_3^n \in\calD_m^c}}  \bbP_V((Y_2^n,Y_3^n) =(y_2^n,y_3^n)| M=m)  \nn\\
 &\qquad\qquad\qquad \times \prod_{x_1, x_2, y_2, y_3} \left(\frac{W(y_2, y_3|x_1, x_2)}{V(y_2, y_3|x_1, x_2)} \right)^{nP(x_1, x_2)U(y_2,y_3|x_1, x_2)} \label{eqn:use_defA}
\end{align}
where \eqref{eqn:change_me} is the key step in this whole proof where we changed the conditional measure (channel) from $W$ to $V$ and  \eqref{eqn:use_defA} follows from the definition of the set $\calA(m, P, U)$ in \eqref{eqn:defB}. Continuing, we have 
\begin{align}
\frac{\rmP_{\rme}(W)}{\rmP_{\rme}(V)} &= \frac{1}{|\calM|}\sum_{m\in\calM} \sum_{ \substack{ P\in\scP_n(\calX_1\times\calX_2) \\  U \in\scV_n(\calY_2\times\calY_3;P)} } \sum_{\substack{ (y_2^n,y_3^n) \in\calA(m, P, U) \\ y_3^n \in\calD_m^c}} \frac{\bbP_V((Y_2^n,Y_3^n) =(y_2^n,y_3^n)| M=m)}{\rmP_{\rme}(V)}  \nn\\
 & \qquad\qquad\qquad \times \exp\left( -n   \sum_{x_1, x_2, y_2, y_3} P(x_1,x_2) U(y_2,y_3|x_1,x_2)\log\frac{V(y_2,y_3|x_1,x_2)}{W(y_2,y_3|x_1,x_2)}\right) \\
 & \ge\exp\Bigg[ -n \bigg(  \frac{1}{|\calM|}\sum_{m\in\calM} \sum_{ \substack{ P\in\scP_n(\calX_1\times\calX_2) \\  U \in\scV_n(\calY_2\times\calY_3;P)} } \sum_{\substack{ (y_2^n,y_3^n) \in\calA(m, P, U) \\ y_3^n \in\calD_m^c}} \frac{\bbP_V((Y_2^n,Y_3^n) =(y_2^n,y_3^n)| M=m)}{\rmP_{\rme}(V)}   \nn\\ 
 & \qquad\qquad\qquad \times     \sum_{x_1, x_2, y_2, y_3} P(x_1,x_2) U(y_2,y_3|x_1,x_2)\log\frac{V(y_2,y_3|x_1,x_2)}{W(y_2,y_3|x_1,x_2)}\bigg)
 \Bigg] \label{eqn:changetoV} .
\end{align}
The last step \eqref{eqn:changetoV}  follows from the convexity  of $t\mapsto \exp(-t)$.  Now the idea is to approximate $U$ with $V$. For this purpose, define the following ``typical'' set 
\begin{equation}
\scG_\gamma(V) \!:= \!\bigg\{ (P,U)  \in\scP(\calX_1\cp\calX_2\cp\calY_2\cp\calY_3) :\!\! \sum_{x_1, x_2, y_2, y_3} P(x_1, x_2) \big|U(y_2,y_3|x_1, x_2) \!-\!V(y_2,y_3|x_1, x_2) \big|\!\le\!\gamma\bigg\}.
\end{equation}
Also define the finite constant 
\begin{equation}
\kappa_V := \max_{x_1, x_2, y_2, y_3: V(y_2,y_3|x_1,x_2) > 0 } - \log V(y_2,y_3|x_1,x_2).
\end{equation}
For $(P,U)\in\scG_\gamma(V)$, it can be  verified by using the definition of $D(V\|W|P)$ \cite[Prop.~13]{Palai11} that 
\begin{equation}
  \sum_{x_1, x_2, y_2, y_3} P(x_1,x_2) U(y_2,y_3|x_1,x_2)\log\frac{V(y_2,y_3|x_1,x_2)}{W(y_2,y_3|x_1,x_2)}\le D(V\|W|P) + \gamma\max\{ \kappa_V,\kappa_W\} .\label{eqn:inJ}
 \end{equation}
 For the typical part of the exponent in~\eqref{eqn:changetoV}, we have 
\begin{align}
T &:= \frac{1}{|\calM|}\sum_{m\in\calM} \sum_{ \substack{ P\in\scP_n(\calX_1\times\calX_2) \\  U \in\scV_n(\calY_2\times\calY_3;P) \\ (P,U) \in\scG_\gamma(V)} } \sum_{\substack{ (y_2^n,y_3^n) \in\calA(m, P, U) \\ y_3^n \in\calD_m^c}} \frac{\bbP_V((Y_2^n,Y_3^n) =(y_2^n,y_3^n)| M=m)}{\rmP_{\rme}(V)}   \nn\\ 
 & \qquad\qquad\qquad \times     \sum_{x_1, x_2, y_2, y_3} P(x_1,x_2) U(y_2,y_3|x_1,x_2)\log\frac{V(y_2,y_3|x_1,x_2)}{W(y_2,y_3|x_1,x_2)} \\
 &\le \left[ \max_P D(V\|W|P) + \gamma\max\{ \kappa_V,\kappa_W\} \right] \nn\\
 & \qquad\qquad\qquad \times   \frac{1}{|\calM|}\sum_{m\in\calM}\sum_{ \substack{ P\in\scP_n(\calX_1\times\calX_2) \\  U \in\scV_n(\calY_2\times\calY_3;P) \\ (P,U) \in\scG_\gamma(V) } } \sum_{\substack{ (y_2^n,y_3^n) \in\calA(m, P, U) \\ y_3^n \in\calD_m^c}} \frac{\bbP_V((Y_2^n,Y_3^n) =(y_2^n,y_3^n)| M=m)}{\rmP_{\rme}(V)}   
 \end{align}  
 Now we drop the condition $(P,U)\in\scG_\gamma(V)$ and continue bounding $T$ as follows
 \begin{align}
 T
  &\le \left[ \max_P D(V\|W|P) + \gamma\max\{ \kappa_V,\kappa_W\} \right] \nn\\
 & \qquad\qquad\qquad \times   \frac{1}{|\calM|}\sum_{m\in\calM}\sum_{ \substack{ P\in\scP_n(\calX_1\times\calX_2) \\  U \in\scV_n(\calY_2\times\calY_3;P) } }  \frac{\bbP_V((Y_2^n,Y_3^n)  \in \calA(m,P,U) , Y_3^n \in \calD_m^c| M=m)}{\rmP_{\rme}(V)}   \\
  & =\left[ \max_P D(V\|W|P) + \gamma\max\{ \kappa_V,\kappa_W\} \right]   \frac{1}{|\calM|}\sum_{m\in\calM} \frac{\bbP_V(Y_3^n\in\calD_m^c | M=m)}{\rmP_{\rme}(V)}   \\
  & =\max_P D(V\|W|P) + \gamma\max\{ \kappa_V,\kappa_W\}.\label{eqn:inJ2}
\end{align}
The last step follows from the definition of the average error probability $\rmP_{\rme}(V)$ in \eqref{eqn:def_err_prob}. 

For $(P,U)\notin\scG_\gamma(V)$, by Pinsker's inequality \cite[Ex.~3.18]{Csi97} and Jensen's inequality (see \cite[Lem.~19]{Palai11}), 
\begin{align}
D(U \| V|P) &\ge\bbE_{P} \left[ \frac{1}{2\ln 2} \| U(\cdot,\cdot|X_1, X_2) - V(\cdot,\cdot|X_1, X_2) \|_1^2\right] \\
 &\ge \frac{1}{2\ln 2}  \left[  \bbE_{P} \left( \big| U(\cdot,\cdot|X_1, X_2) - V(\cdot,\cdot|X_1, X_2)  \big|\right)\right]^2  \ge\frac{\gamma^2}{2\ln 2}.\label{eqn:div_bd}
\end{align}
 Furthermore, for any $(y_2^n,y_3^n) \in\calA(m,P,U)$, 
\begin{align}
&\log \bbP_V((Y_2^n,Y_3^n) =  (y_2^n,y_3^n)| M=m)  \nn\\*
&= \sum_{i=1}^n\log V(y_{2i}, y_{3i} | f_i(m), g_i(y_2^{i-1} ) )  \\
&= n\sum_{x_1, x_2, y_2, y_3} \left(\frac{1}{n}\sum_{i=1}^n \bone\{f_i(m)=x_1, g_i(y_2^{i-1} ) =x_2,  y_{2i}=y_2, y_{3i}=y_3 \}\right)  \log V(y_{2 }, y_{3 } | x_1, x_2 )    \\
&= n\sum_{x_1, x_2, y_2, y_3} P(x_1, x_2) U(y_2, y_3|x_1, x_2) \log V(y_{2 }, y_{3 } | x_1, x_2 )    \label{eqn:grouping}\\
 & = -n (D(U \| V|P ) + H(U|P)) ,
\end{align} 
where \eqref{eqn:grouping} follows from the definition of the set $\calA(m,P,U)$.
So in a similar manner as in \cite[Prop.~14(c)]{Palai11}, we have  $|\calA(m,P,U)|\le\exp(n H(U|P))$. Thus, 
\begin{align}
&\sum_{  \substack{  P\in\scP_n(\calX_1\times\calX_2) \\  U \in\scV_n(\calY_2\times\calY_3;P)  \\ (P,U) \notin\scG_\gamma(V) }} \bbP_V((Y_2^n,Y_3^n)\in \calA(m,P,U) | M=m) \nn\\
 &=\sum_{  \substack{  P\in\scP_n(\calX_1\times\calX_2) \\  U \in\scV_n(\calY_2\times\calY_3;P)  \\ (P,U) \notin\scG_\gamma(V) }} | \calA(m,P,U)|\exp[-n (D(U \| V|P ) + H(U|P))]  \\
&\le \sum_{  \substack{  P\in\scP_n(\calX_1\times\calX_2) \\  U \in\scV_n(\calY_2\times\calY_3;P)  \\ (P,U) \notin\scG_\gamma(V) }} \exp(-n D(U\| V|P)) \\
&\le  (n+1)^{|\calX_1| |\calX_2| |\calY_2| |\calY_3| } \exp\left(-n\frac{\gamma^2}{2\ln 2} \right) , \label{eqn:notinJ}
\end{align}
where the final step follows from  \eqref{eqn:div_bd} and  Lemma \ref{lem:types}.
As a result,  for the atypical part of the exponent in~\eqref{eqn:changetoV},
\begin{align}
S &:= \frac{1}{|\calM|}\sum_{m\in\calM} \sum_{ \substack{ P\in\scP_n(\calX_1\times\calX_2) \\  U \in\scV_n(\calY_2\times\calY_3;P) \\ (P,U) \notin\scG_\gamma(V)} } \sum_{\substack{ (y_2^n,y_3^n) \in\calA(m, P, U) \\ y_3^n \in\calD_m^c}} \frac{\bbP_V((Y_2^n,Y_3^n) =(y_2^n,y_3^n)| M=m)}{\rmP_{\rme}(V)}   \nn\\ 
 & \qquad\qquad\qquad \times     \sum_{x_1, x_2, y_2, y_3} P(x_1,x_2) U(y_2,y_3|x_1,x_2)\log\frac{V(y_2,y_3|x_1,x_2)}{W(y_2,y_3|x_1,x_2)} \\
 &\le  \frac{\kappa_W }{\rmP_{\rme}(V)}\cdot \frac{1}{|\calM|}\sum_{m\in\calM} \sum_{ \substack{ P\in\scP_n(\calX_1\times\calX_2) \\  U \in\scV_n(\calY_2\times\calY_3;P) \\ (P,U) \notin\scG_\gamma(V)} }  \bbP_V((Y_2^n,Y_3^n)\in \calA(m,P,U) | M=m)  \label{eqn:before_notinJ2} \\
& \le\frac{\kappa_W }{\rmP_{\rme}(V) }  (n+1)^{|\calX_1| |\calX_2| |\calY_2| |\calY_3| } \exp\left(-n\frac{\gamma^2}{2\ln 2} \right).\label{eqn:notinJ2}
\end{align}
In \eqref{eqn:before_notinJ2} we upper bounded $V(y_2,y_3|x_1,x_2)$ by $1$ and $-\log W(y_2,y_3|x_1,x_2)$ by $\kappa_W$ for those $(x_1, x_2, y_2, y_3)$ such that $W(y_2,y_3|x_1,x_2)>0$. If instead $W(y_2,y_3|x_1,x_2)=0$ and $U(y_2,y_3|x_1,x_2)>0$ for some $(y_2^n,y_3^n)$ in the sum in $S$, the probability $\bbP_V((Y_2^n,Y_3^n) =(y_2^n,y_3^n)| M=m)=0$ so these symbols can be ignored.  Combining \eqref{eqn:changetoV}, \eqref{eqn:inJ2} and \eqref{eqn:notinJ2}, we conclude that 
\begin{equation}
\frac{\rmP_{\rme}(W)}{\rmP_{\rme}(V)}\ge\exp[-n(T+S)]\ge \exp\left[-n \left( \max_P D(V\|W|P) + \varrho_{n,\gamma}\right) \right] \label{eqn:conclude}
\end{equation}
where  
\begin{equation}
\varrho_{n,\gamma}:=\gamma\max\{ \kappa_V,\kappa_W\}+\frac{\kappa_W }{\rmP_{\rme}(V) }  (n+1)^{|\calX_1| |\calX_2| |\calY_2| |\calY_3| } \exp\left(-n\frac{\gamma^2}{2\ln 2} \right).
\end{equation}
Now assume that  the DM-RC $V$ is chosen to achieve  the minimum in \eqref{eqn:ub_rel} evaluated at $R-\delta$, i.e., $V \in \argmin_{V: C_{\mathrm{cs}}(V)\le R-\delta}\max_{P_{X_1 X_2}}D(V\| W|P_{X_1 X_2})$. Then uniting~\eqref{eqn:Vge} and \eqref{eqn:conclude} yields
\begin{align}
\rmP_{\rme}(W)\ge \eta\exp\left[-n \left( E_{\mathrm{cs}}(R-\delta) + \varrho_{n,\gamma}\right) \right].
\end{align}
We then obtain
\begin{equation}
 \frac{1}{n}\log \frac{1}{ \rmP_{\rme}(W) } \le E_{\mathrm{cs}}(R-\delta) + \varrho_{n,\gamma}-\frac{\log \eta}{n}.
\end{equation}
Let $\gamma=n^{-1/4}$ so $\varrho_{n,\gamma}\to 0$ as $n\to\infty$. Note also that $\eta>0$. As such by taking limits,
\begin{equation}
\liminf_{n\to\infty}  \frac{1}{n}\log \frac{1}{ \rmP_{\rme}(W) } \le E_{\mathrm{cs}}(R-\delta) .
\end{equation}
Since the left-hand-side does not depend on $\delta$, we may now take the limit of the right-hand-side as $\delta\to 0$ and use the continuity of $E_{\mathrm{cs}}(R)$ (which follows the continuity of $V\mapsto \rvC_{\mathrm{cs}}(V)$ and $V\mapsto\max_{P_{X_1 X_2}}D(V\| W| P_{X_1 X_2})$) and obtain 
\begin{equation}
\liminf_{n\to\infty}  \frac{1}{n}\log \frac{1}{ \rmP_{\rme}(W) } \le E_{\mathrm{cs}}(R ) 
\end{equation}
as desired.
 \end{proof}

\section{Numerical Evaluation for the Sato Relay Channel} \label{sec:num}

In this section we evaluate the  error exponent for the PDF (or, in this case, decode-forward) scheme  presented in Theorem~\ref{thm:pdf} for a canonical DM-RC, namely the Sato relay channel~\cite{sato76}. This is a physically degraded relay channel in which $\calX_1=\calY_2=\calY_3=\{0,1,2\}$, $\calX_2=\{0,1\}$, $Y_2=X_1$ (deterministically) and the transition matrices  from $(X_1, X_2)$ to $Y_3$ are
\begin{align}
W_{Y_3|X_1, X_2 }(y_3|x_1 , 0) & = \begin{bmatrix}
1 & 0 & 0  \\ 0  & 0.5 & 0.5 \\    0  & 0.5 & 0.5 
\end{bmatrix} ,  \\
 W_{Y_3|X_1, X_2}(y_3|x_1 , 1) & = \begin{bmatrix}
0  & 0.5 & 0.5 \\    0  & 0.5 & 0.5  \\ 1 & 0 & 0  
\end{bmatrix}  .
\end{align}
It is known that the capacity of the Sato relay channel is $C_{\mathrm{Sato}}=1.161878$ bits per channel use~\cite[Eg.~16.1]{elgamal} and the capacity-achieving input distribution is 
\begin{equation}
Q_{X_1   X_2}^* (x_1, x_2) = \begin{bmatrix}
p & q  \\
q& q\\
 q & p  
\end{bmatrix} \label{eqn:caid}
 \end{equation} 
where $p = 0.35431$ and $q= 0.072845$ \cite[Table I]{CEG}. The auxiliary random variable $U$ is set to $X_1$ in this case. It is easy to check that $I(X_1X_2;Y_3)=I(X_1;Y_2 |X_2)=C_{\mathrm{Sato}}$ ($I(X_1X_2;Y_3)$ and $ I(X_1;Y_2 |X_2)$ are the only two relevant mutual information terms in the DF lower bound and the cutset upper bound) when the distribution of $(X_1,X_2)$  is  capacity-achieving, i.e., the random variables   $(X_1,X_2)$ have distribution $Q_{X_1X_2}^*$ in \eqref{eqn:caid}.  We study the effect of the number of blocks in block-Markov coding in the following. We note that this is the first numerical study of error exponents on a discrete relay channel. Other numerical  studies were done for  continuous-alphabet relay channels, e.g.,~\cite{Ngo10,Bra12, LiGeor,zhangmitra,WenBerry,Yilmaz}.

\begin{figure}
\centering
\includegraphics[width = .95\columnwidth]{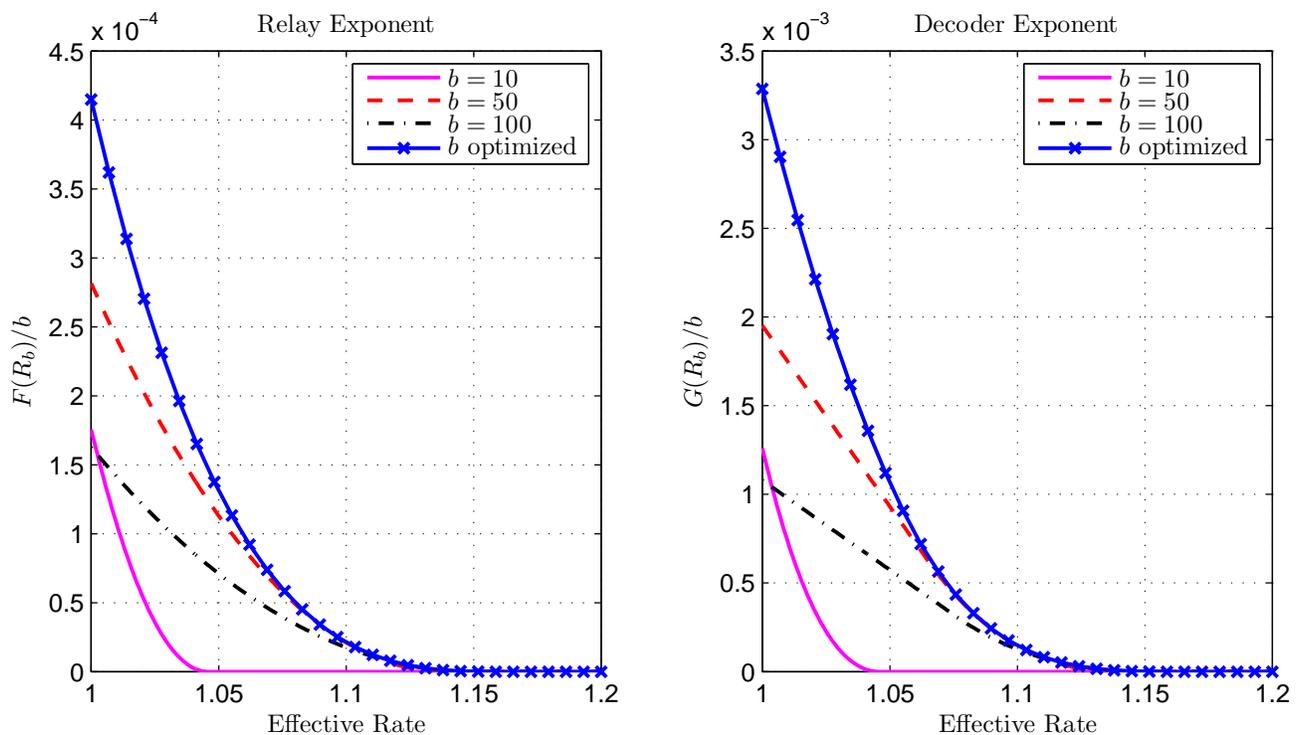}
\caption{Plots of the  relay and decoder exponents divided by the number of blocks  $b$ against effective rate $\Reff$ for the Sato relay  channel. The optimized exponents $\Reff\mapsto \max_b F(R_b)/b$ and $\Reff\mapsto \max_b G(R_b)/b$ are also shown. }
\label{fig:sato}
\end{figure}

We set the number of blocks $b\in \{10,50,100\}$. In view that the capacity is  $C_{\mathrm{Sato}}=1.161878$, we   let effective rate $\Reff$, defined in \eqref{eqn:reff}, be in the range $[1.00,1.20]$ bits per channel use. The {\em per-block rate}  can thus be computed as
\begin{equation}
R_b :=  \frac{b}{b-1}\Reff
\end{equation}
where we regard $\Reff$ as fixed and constant and made the dependence of the per-block rate on $b$ explicit.
Since error exponents are monotonically decreasing in the rate (if the rate is smaller than $C_{\mathrm{Sato}}$),   
\begin{equation}
b_1<b_2\,\,\Rightarrow\,\,  F(R_{b_1})<F(R_{b_2}) \label{eqn:degrade_F}
\end{equation}  
because $R_{b_1}>R_{b_2}$  and similarly for $G(R_b)$. We reiterate that $\tilG(R_b'')$, defined in \eqref{eqn:G2}, is not relevant for the numerical evaluation for this example. Note  that there is a tradeoff concerning $b$ here. If $b$ is small, the degradation of the exponent due to the effect in~\eqref{eqn:degrade_F} is significant but we divide $F(R_b)$ by a smaller factor in~\eqref{eqn:pdf_lb}. Conversely, if $b$ is larger, the degradation due to \eqref{eqn:degrade_F} is negligible but we divide by a larger number of blocks to obtain  the overall exponent. We evaluated the  {\em relay exponent}  $F(R_b)$ and the {\em decoder exponent} $G(R_b)$ in their Gallager forms in \eqref{eqn:gal1} and \eqref{eqn:gal2} using the capacity-achieving input distribution in~\eqref{eqn:caid}.   In Fig.~\ref{fig:sato}, we plot the exponents $F(R_b)$ and $G(R_b)$ divided by $b$  as   functions of the effective rate $\Reff$. For each $\Reff$, we also optimized for the largest exponents over $b$ in both cases.

We make a three observations concerning Fig.~\ref{fig:sato}. 

\begin{enumerate}

\item  First, the relay exponent $F(R_b)$ dominates because it is uniformly smaller than the decoder exponent $G(R_b)$. Hence, the overall exponent for the Sato channel using decode-forward is the relay exponent (the scales on the vertical axes are different). Since we made this observation, we also evaluated the PDF exponent with a non-trivial $U$ (i.e., not equal to $X_1$) whose alphabet $|\calU|$ was allowed to be as large as $10$,  non-trivial split of $R$ into $R'$ and $R''$, while preserving $Q_{X_1 X_2}^*$ in \eqref{eqn:caid}  as the input distribution. This was done   to possibly increase the overall (minimum) exponent in \eqref{eqn:gal1}--\eqref{eqn:gal3}. However, from our numerical studies, it appears that there was no advantage in using PDF in the error exponents sense for the Sato  relay channel.   

\item Second, the cutoff rates can be seen to be $C_{\mathrm{Sato}}$ bits per channel use in both plots and this can only be  achieved by letting $b$ become large. This is consistent with block-Markov coding \cite[Sec.~16.4.1]{elgamal} in which to achieve the DF lower bound asymptotically, we need to let $b$ tend to infinity in addition to letting the per-block blocklength $n$ tend to infinity.

\begin{figure}
\centering
\includegraphics[width = .95\columnwidth]{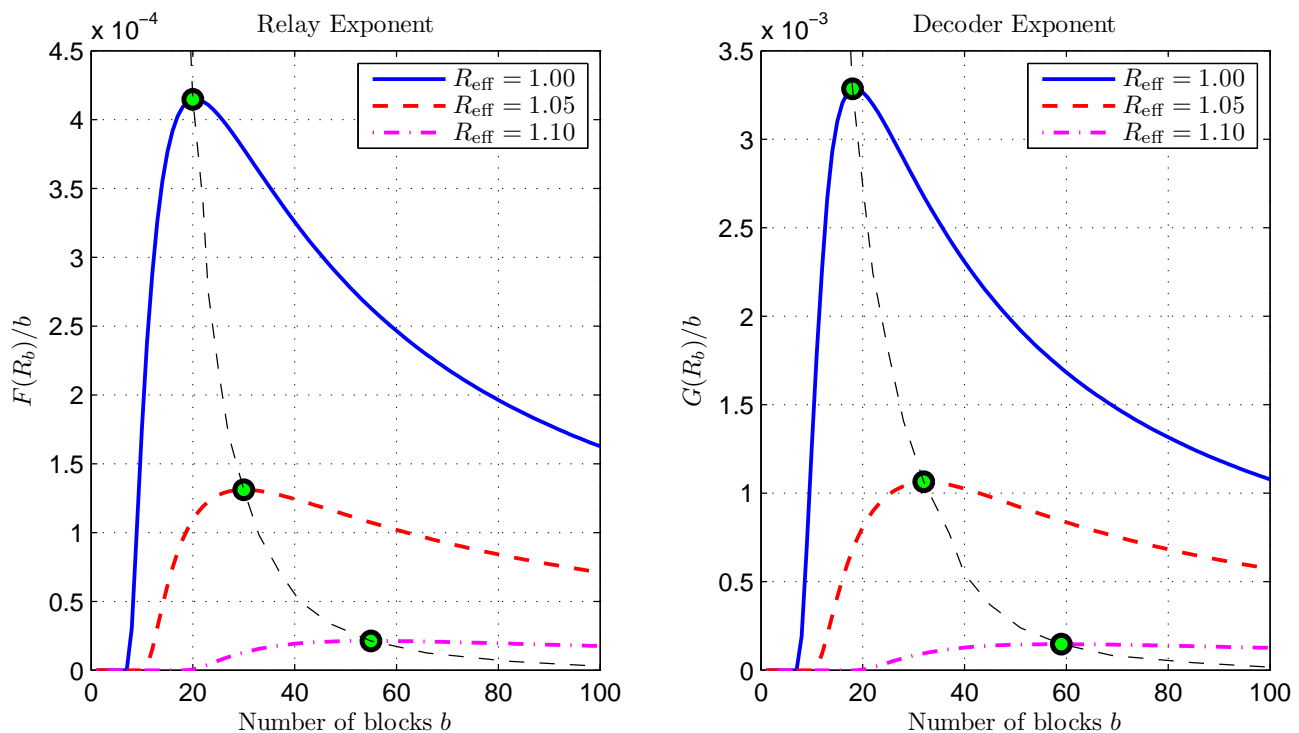}
\caption{Plots of the relay and decoder exponents divided by the number of blocks $b$  against $b$  for the Sato relay  channel for $\Reff \in \{1.00, 1.05, 1.10\}$ bits per channel use. The green circles represent the maxima of the curves and shows that the optimum number of blocks increases as the effective rate one wishes to operate at increases. The broken black line traces the path of the optimal number of blocks as $\Reff$ increases. }
\label{fig:exp_blocks}
\end{figure}

\item Finally, observe that  if one would like to operate at an effective   rate  slightly below $\Reff=1.10$    bits per channel use (say), we would choose the number of blocks $b\in \{10,50,100\}$ to be a moderate $50$ instead of the larger $100$ so as to attain a larger overall exponent. This is because the overall exponent in~\eqref{eqn:pdf_lb} is the ratio of the relay exponent $F(R_b)$ and the number of blocks $b$ and dividing by $b$ degrades the overall exponent more than the effect in \eqref{eqn:degrade_F}. This implies that to achieve a small error probability at a fixed rate, we should use a block-Markov scheme with a carefully chosen number of blocks {\em not tending to infinity} assuming the number of channel uses within each block, i.e., the per-block blocklength, is sufficiently large (so that pre-exponential factors are negligible).  However, if we want to operate at rates close to capacity, naturally, we need a larger number of blocks otherwise the exponents are zero.  For example if we   use only $10$ blocks, both exponents are $0$ for all $\Reff>1.05$ bits per channel use. The functions of $b\mapsto F(R_b)/b$ and $b\mapsto G(R_b)/b$  are illustrated in Fig.~\ref{fig:exp_blocks} for $\Reff \in \{1.00 , 1.05, 1.10\}$. 

\end{enumerate}
\section{Conclusion and Future Work}\label{sec:concl}

In this paper, we derived   achievable error exponents for the DM-RC by carefully analyzing PDF and CF. One of the take-home messages is that to achieve the best error decay for a fixed rate, we need to choose a moderate number of blocks in block-Markov coding. We also derived an upper bound for the reliability function by appealing to Haroutunian's techniques for channels with feedback.   
 
It is the author's hope that the present paper will precipitate  research in exponential error bounds for noisy networks with relays. We   discuss a few avenues for further research here. 
\begin{enumerate}
\item  Most importantly,  it is imperative to develop alternate forms of  or approximations to the CF exponent (Theorem~\ref{thm:cf}) and the upper bound on the reliability function (Theorem~\ref{thm:upper}) which are   difficult to compute in their current  forms.  

\item \label{item:ens_typ} It would  also be useful to show  that the error exponents we derived in achievability theorems are {\em ensemble tight}, i.e.,    given the random codebook, the error exponents cannot be improved by using other decoding rules. This establishes  some form of optimality of our coding schemes and analyses.  This was done for the point-to-point channel in~\cite{gallager73} and mismatched decoding in~\cite{Sca12b}. 

\item \label{item:continuous} The results contained herein hinge on the method of types which is well-suited to analyze discrete-alphabet systems. It is also essential, for real-world wireless communication networks, to develop parallels of the main results for continuous-alphabet  systems such as the AWGN relay channel. While there has been some work on this  in \cite{Ngo10,Yilmaz,Bra12},   schemes  such as CF have remained relatively unexplored. However, one needs a Marton-like exponent~\cite{Marton74} for lossy source coding with uncountable alphabets. Such an exponent was derived by Ihara and Kubo~\cite{Ihara00} for Gaussian sources using geometrical arguments. Incorporating Ihara and Kubo's derivations into a CF exponent analysis would be interesting and challenging. 

 \item Since {\em noisy network coding}~\cite{Lim11} is a variant of CF that generalizes various network coding scenarios, in the   future, we hope to also derive an achievable error exponent based on the noisy network coding strategy and compare that to the CF exponent we derived in Theorem~\ref{thm:cf}. In particular, it would be useful to observe  if the resulting noisy network coding exponent is  easier to compute compared to the CF exponent. 
 \item In addition, a combination of DF and CF was used for relay networks with at least $4$ nodes in Kramer-Gastpar-Gupta~\cite{KGG05}. It may be insightful to derive the corresponding error exponents at least for DF and understand how the exponents scale with the number of nodes in a network. 
 \item It is natural to wonder whether the technique presented in Section~\ref{sec:sp} applies to  discrete memoryless relay channels with various forms of feedback~\cite[Sec.~17.4]{elgamal} since techniques from channel  coding with feedback\cite{Har77, Palai11} were employed to derived the upper bound on the reliability function.
 
 \item Finally, we also expect that  the {\em moments of type class enumerator} method by Merhav~\cite{Merhav09, merhav10} and co-authors may yield an alternate forms of the random coding and expurgated exponents that may have a different interpretation (perhaps, from the  statistical physics perspective) vis-\`a-vis the types-based random coding error exponent presented in Section~\ref{sec:dec}. 
\end{enumerate} 


\appendices 
\section{Proof of Lemma~\ref{lem:joint_typ}}\label{app:joint_typ}
\begin{proof}
Because $X_2^n$ is generated uniformly at random from $\calT_V(x_1^n)$, 
\begin{align}
\bbP\left[\bary^n \in \calT_{V'}(x_1^n, X_2^n) \right]=\sum_{x_2^n \in \calT_V(x_1^n) }\frac{1}{|\calT_V(x_1^n)|}\bone\{\bary^n \in \calT_{V'}(x_1^n, x_2^n) \}. \label{eqn:proof_lem_jt}
\end{align}
Consider  reverse channels $\tilV:\calX_1\times\calY\to\calX_2$ and let $\scR(V)$ the be collection  of reverse channels satisfying $\sum_{y}\tilV(x_2|x_1,y)W(y|x_1)=V(x_2|x_1)$ for all $x_1, x_2, y$.  Note that $\bary^n \in \calT_{V'}(x_1^n, x_2^n)$ holds if and only if there exists some $\tilV\in\scV_n(\calX_2; P\times W)\cap\scR(V)$  such that $x_2^n \in \calT_{\tilV}(x_1^n, \bary^n)$. Then   we may rewrite~\eqref{eqn:proof_lem_jt} as 
\begin{align}
\bbP\left[\bary^n \in \calT_{V'}(x_1^n, X_2^n) \right] &=\sum_{\tilV\in\scV_n(\calX_2; P\times W)\cap\scR(V)}\frac{ | \calT_{\tilV}(x_1^n, \bary^n) |}{|\calT_V(x_1^n)|}  \\
&\le\sum_{\tilV\in\scV_n(\calX_2; P\times W)\cap\scR(V)} \frac{\exp(n H(\tilV|P\times W))}{(n+1)^{ -  |\calX_1||\calX_2|}\exp(n H(V|P))} \\
&\le (n+1)^{ | \calX_1 | |\calX_2 | ( |\calY| + 1) }\frac{\exp(n H(\tilV^*|P\times W))}{ \exp(n H(V|P))}  \label{eqn:largest_V}\\
&= p_2(n)  \exp[-nI(V, V'|P)], \label{eqn:largest_V2}
\end{align}
where in \eqref{eqn:largest_V}, $\tilV^*\in\scV_n(\calX_2; P\times W)$ is the conditional type that maximizes the conditional entropy $H(\tilV|P\times W)$ subject to the constraint that it also belongs to $\scR(V)$; and in \eqref{eqn:largest_V2},   $p_2(n)$ is some polynomial function of $n$ given in the previous expression, and the equality follows from the fact that $I(X_2;Y|X_1)=H(X_2|X_1)-H(X_2|X_1 Y)$ and marginal consistency.  The lower bound proceeds similarly, 
\begin{align}
\bbP\left[\bary^n \in \calT_{V'}(x_1^n, X_2^n) \right] \ge \frac{(n+1)^{ -  |\calX_1||\calX_2||\calY|} \exp(n H(\tilV^*|P\times W))}{\exp(n H(V|P))}=  \frac{1}{p_1(n)} \exp[-nI(V, V'|P)],
\end{align}
where $p_1(n)$ is some polynomial. This proves the lemma. 
\end{proof}

\section{Proof of Lemma~\ref{lem:joint_typ2}}\label{app:joint_typ2}

\begin{proof}
Because  $X_2^n$ is uniformly distributed in $\calT_V(x_1^n)$, we have
\begin{equation}
\bbP\left[X_2^n \in \calT_W(y^n, x_1^n)\right]=\sum_{ x_2^n \in\calT_V(x_1^n)}\frac{1}{ |\calT_V(x_1^n)|} \bone\{x_2^n\in\calT_W(y^n, x_1^n)\}.
\end{equation}
As a result,
 \begin{equation}
\bbP\left[X_2^n \in \calT_W(y^n, x_1^n)\right]= \frac{ |\calT_V(x_1^n)\cap\calT_W(y^n, x_1^n)|}{ |\calT_V(x_1^n)|}\le\frac{ | \calT_W(y^n, x_1^n)|}{ |\calT_V(x_1^n)|}\le\frac{\exp(nH(W|P\times V')}{ (n+1)^{-|\calX_1| |\calX_2|}\exp(nH(V|P))}.
\end{equation}
 Thus, denoting $p_3(n)$ as some polynomial function of $n$, we have
 \begin{equation}
\bbP\left[X_2^n \in \calT_W(y^n, x_1^n)\right]\le p_3(n)\exp[-n I(V', W|P)]
\end{equation}
because $W$  satisfies the marginal consistency property in the statement of the lemma and $I(X_2;Y|X_1)=H(X_2|X_1)-H(X_2|X_1 Y)$.
\end{proof}

\subsubsection*{Acknowledgements} I am extremely grateful to Yeow-Khiang Chia and Jonathan Scarlett for many helpful discussions and comments that helped to improve the content and the  presentation in this work. 
I  would also like to sincerely acknowledge the Associate Editor Aaron Wagner and the two  anonymous reviewers for their extensive and useful comments during the revision process.

\bibliographystyle{IEEEtran}
\bibliography{isitbib}

\begin{IEEEbiographynophoto}{Vincent Y. F. Tan} (S'07-M'11)  is an Assistant Professor   in the Department of Electrical and Computer Engineering (ECE) and the Department of  Mathematics at the National University of Singapore (NUS).  He received the B.A.\ and M.Eng.\ degrees in Electrical and Information Sciences from  Cambridge University in 2005. He received the Ph.D.\ degree in Electrical Engineering and Computer Science (EECS) from the Massachusetts Institute of Technology in 2011. He was   a postdoctoral researcher in the Department of  ECE  at the University of Wisconsin-Madison and following that, a research scientist at the Institute for Infocomm (I$^2$R) Research,  A*STAR, Singapore. His research interests include information theory, machine learning and signal processing.

Dr.\ Tan   received the  MIT EECS Jin-Au Kong outstanding doctoral thesis prize in 2011 and the NUS Young Investigator Award in 2014.  He has authored a research monograph on {\em Asymptotic Estimates in Information Theory with Non-Vanishing Error Probabilities} in the Foundations and Trends\textsuperscript{\textregistered} in Communications and Information Theory Series (NOW Publishers).
\end{IEEEbiographynophoto}

\end{document}